\documentclass[aip,superscriptaddress,jcp,showpacs,floatfix,preprint]{revtex4}
\usepackage{epsfig}
\usepackage{graphicx}
\usepackage{amssymb}
\usepackage{amsmath}
\usepackage{epstopdf}

\begin{document}

\title{
Self-assembly of bi-functional patchy particles with anisotropic shape into polymers chains: 
theory and simulations}
\author{Cristiano~De Michele}
\affiliation{Dipartimento di Fisica,
         Universit\`a di Roma {\em La Sapienza}, P.le A. Moro 2, 00185 Roma, Italy}
\author{Tommaso Bellini}
\affiliation{Universit\`a degli studi di Milano,  Dipartimento Chim. Biochim. \& Biotecnol. Med., I-20090 Milan, Italy }
\author{Francesco~Sciortino} 
\affiliation{Dipartimento di Fisica  and  CNR-ISC,
         Universit\`a di Roma {\em La Sapienza}, P.le A. Moro 2, 00185 Roma, Italy}
\date{\today}
\begin{abstract}
Concentrated solutions of short blunt-ended DNA duplexes, down to 6 base pairs, are known to order into the nematic liquid crystal phase.  This self-assembly is due to the stacking interactions between the duplex terminals that promotes their aggregation into poly-disperse chains with a significant persistence length. Experiments show that liquid crystals phases form above a critical volume fraction depending on the duplex length. 
We introduce and investigate via numerical simulations, a coarse-grained model of DNA double-helical duplexes. Each duplex is represented as an hard quasi-cylinder whose bases are decorated with two identical reactive sites.
The stacking interaction between terminal sites is modeled via a short-range square-well potential. 
We compare the numerical results with predictions based on a free energy functional and find satisfactory quantitative matching of the isotropic-nematic phase boundary and of the system structure. 
Comparison of numerical and theoretical results with experimental findings 
confirm that the DNA duplexes self-assembly can be properly modeled via equilibrium polymerization of cylindrical particles and enables us to estimate the stacking energy. 

\end{abstract}
\maketitle
\section{Introduction}
Self-assembly is the spontaneous organization of matter into reversibly-bound aggregates.  
In contrast to chemical synthesis where molecular complexity is achieved through covalent bonds, in self-assembled structures the
molecules or supramolecular aggregates spontaneously form following the minimization of their free energy. 
Self-assembly is ubiquitous in nature and can involve the structuring of elementary building blocks of various sizes, ranging from 
simple molecules (e.g. surfactants) to the mesoscopic units (e.g. colloidal particles), thus being of topical interest in several fields, including soft matter and biophysics \cite{HamleyBook,Glotz_04,WhitesidesPNAS02}.  
Understanding and thus controlling the processes of self-assembly is important in material science and technology
for devicing new materials whose physical properties are controlled by tuning the interactions of the assembled components
\cite{Workum_06,MIRKIN_96,Manoh_03,Cho_05,Yi_04,Cho_05,starrnano,dnastarr,Stupp_93,doye}.

A particular but very interesting case of self-assembly occurs when the anisotropy of attractive interactions
between the monomers favors the formation of linear or filamentous aggregates, i.e. linear semi-flexible, flexible or rigid chains.
A longstanding example is provided by the formation of worm-like micelles of amphiphilic molecules in water or  microemulsions of water and oil which are stabilized by amphiphilic molecules . If supramolecular aggregates possess a sufficient rigidity the system may exhibit liquid crystal (LC) ordering even if the self-assembling components do not have the required shape anisotropy to guarantee the formation of
nematic phases. An intense experimental activity has been dedicated to the study of nematic transitions in micellar 
systems \cite{Khan96,CatesLangmuir94,KnutzSM08}.
Another prominent case is that of formation of fibers and fibrils of peptides and proteins \cite{MezzengaLangmuir2010,LeePRE09,CiferriLC07,AggeliJACS03}.
Over last $50$ years LC phases have been also observed in solutions of long duplex B-form DNA composed of $10^2$ to $10^6$ base pairs  \cite{RobinsonTetra61,LivolantNature89,MerchantBJ97,FerrariniJCP05} and in the analogous case of filamentous viruses \cite{FerrariniPRL06,DogicSM09,FredenPRL03,TomarJACS07,MinskyNatCell02}.
More recently, a series of experiments\cite{BelliniScience07,BelliniJPCM08,BelliniPNAS2010},
have provided evidence that also a solution of short DNA duplexes (DNAD), 6 to 20 base pairs in length can form liquid crystal  ordering
above a critical concentration, giving rise to  nematic and 
columnar LC phases\cite{BelliniScience07}.

\begin{figure}[tbh]
\vskip 0.2cm
\includegraphics[width=.49\textwidth]{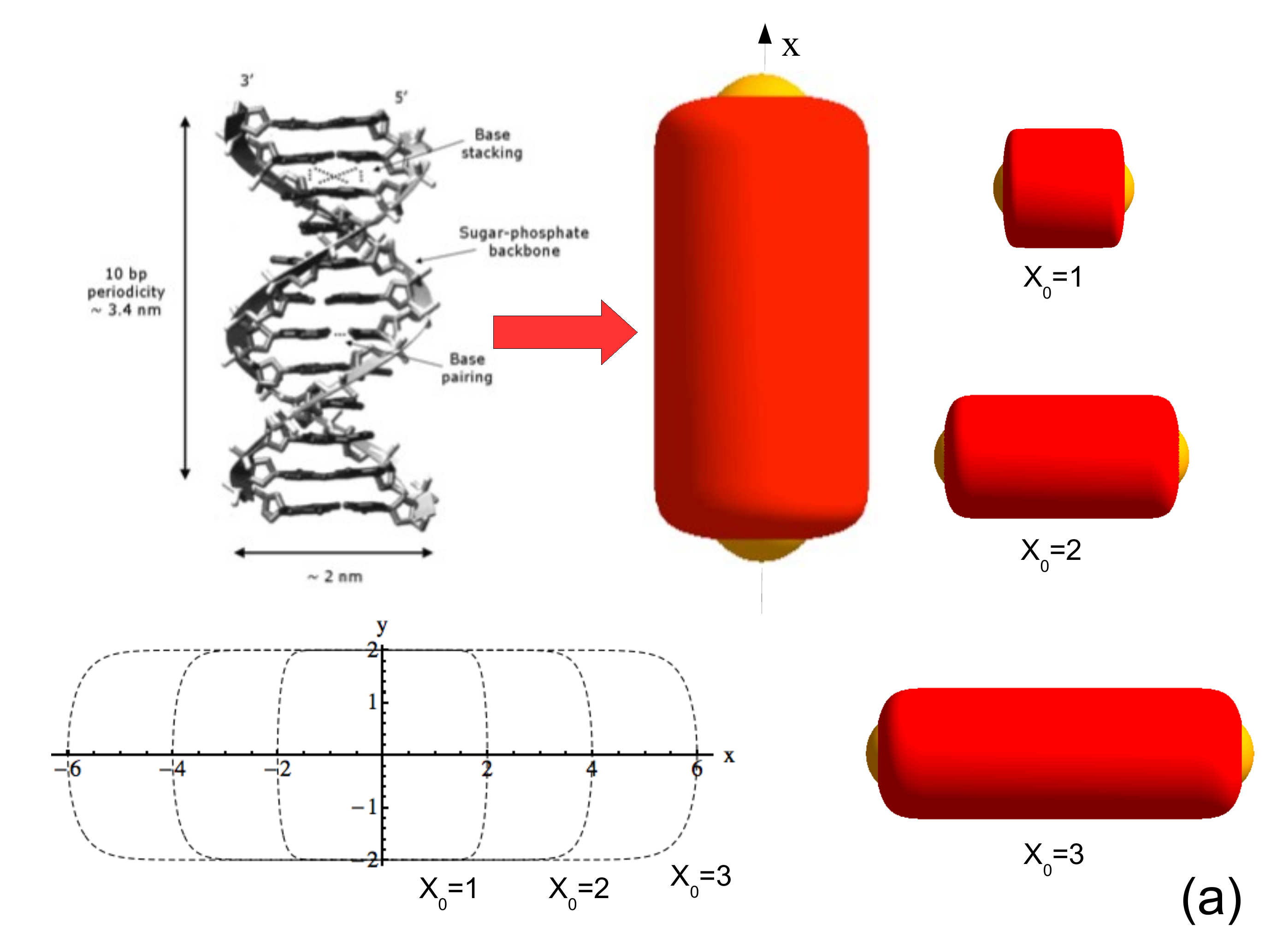}
\includegraphics[width=.49\textwidth]{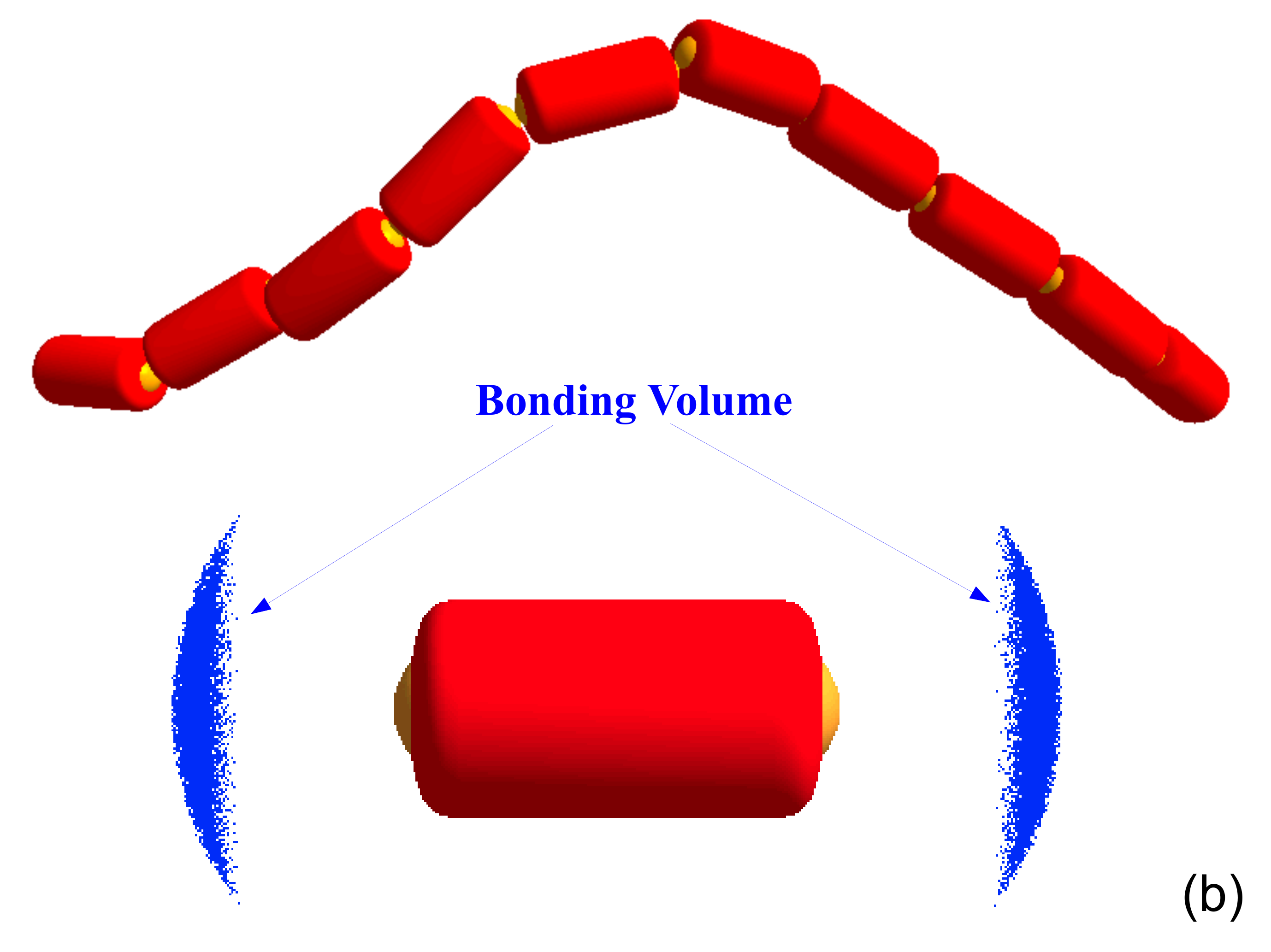}
\caption{Coarse-grained model of DNA duplexes. (a)  DNA duplex and a 3D graphical representation of its corresponding coarse-grained model comprising a SQ, symmetric around the $x$ axis, decorated with two sticky spots located on its bases.  The figure also show SQs of different aspect ratios ($X_0=1,2,3$) and 
the projection of their surfaces onto the $xy$-plane.   Note that the base roundness increases on increasing
$X_0$.
(b) A random chain of $10$ monomers and a representation (blue clouds) of the  points where
the  center of mass of a different monomer can be located in a bonding configuration. This set of points
defines the  bonding volume.}
\label{Fig:singleSQ}
\end{figure}

%
This behavior was found when the terminals of the duplexes interact attractively. This condition is verified either when duplexes terminate bluntly, as in the case of fully complementary strands shown in Fig. \ref{Fig:singleSQ}a, or when the strands arrange in shifted double-helices whose overhangs are mutually interacting. This behavior is not restricted to B-form DNA oligomers, as it has also been observed in solutions of blunt-ended A-form RNA oligomeric duplexes \cite{BelliniJACS08}. As terminals are modified to disrupt attraction, the LC long range ordering is lost. 
Overall, the whole body of experimental evidence supports the notion that LC formation is due to the formation of reversible linear aggregates of duplexes, in turn promoting the onset of long-ranged LC orientational ordering. According to this picture, the LC ordering of oligomeric DNA is analogous to the LC ordering of chromonic liquid crystals \cite{LydonJMC10}. Both in chromonics and in blunt-ended DNA duplexes, the aggregation takes place because of stacking interaction, generally understood as hydrophobic forces acting between the flat hydrocarbon surfaces provided by the core of chromonic molecules and by the paired nucleobases at the duplex terminals \cite{KoolJACS00, BelliniReview2011}. 

The LC ordering of nucleic acids is relevant for various reasons. Firstly, it provides a new model of reversible aggregation leading to macroscopic ordering in which the strength of the inter-monomer attraction can be modified by changing the duplex terminals (bunt-end stacking or pairing of overhangs). Second, it provides a new access to the DNA-DNA interactions, and in particular to stacking interactions, whose nature is still investigated and debated \cite{KoolJACS00,BelliniReview2011}. In this vein, self-assembly acts as an amplifier of the inter-monomeric interactions, enabling studying the effects of minor molecular modification (e.g. oligomer terminations) on the base stacking.  Finally, stacking and self-assembly are often invoked as the prebiotic route to explain the gap between the random synthesis of elementary carbon-based molecules
and the first complex molecules, possibly RNA oligomers, capable of catalyze their own synthesis \cite{SzostakARB10}.
To proceed in any of these directions, it is necessary to rely on models enabling to quantitatively connect the collective behavior of nucleic acids oligomers to the molecular properties, and in particular to the duplex size and to the strength and range of the interduplex attractions.

While the nematization transition in rigid and semiflexible polymers has been investigated in details in the
past and rather accurate thermodynamic descriptions have been proposed\cite{Vroege92,DijkstraPRE97,Semenov81,Semenov82,MulderJPCM06,DijkstraPRL11,WesselsSM03,ChenMacromol93,Odijk86}, much less is known for the case in which the nematic transition takes place in equilibrium polymer systems, i.e. when
the average length of the chains depends on the state point explored.
Recent theoretical and numerical works \cite{GlaserMC,KindtJCP04}
has renewed the interest in this topic\cite{CatesEPL94}. Ref.~\cite{KindtJCP04} investigate the self-assembly
and nematization of spheres, while Ref.~\cite{GlaserMC} focuses on polymerization of
interacting cylinders.  In this article we propose  a coarse-grained model similar to the one introduced in \cite{GlaserMC}
 and devised to capture the essential physical features of equilibrium polymerization of DNA duplexes
and study it numerically via Monte Carlo simulations  in the constant temperature and pressure ensembles, applying special biasing technique\cite{SiepmannJPCB00,SiepmannJPCB01} to speed up  the equilibration process. 
We then develop a free-energy functional,  building on Wertheim \cite{WertheimJSP1,WertheimJSP2,WertheimJSP3} and Onsager \cite{Onsager49} theories  which provides, 
a satisfactory description of the system in the isotropic  and nematic phases.
A comparison of the calculated phase boundaries for different aspect ratio and different interaction
strength with the experimental results allow us to confirm that the DNAD aggregation and LC ordering processes  can be properly modeled via equilibrium polymerization of  cylindrical particles and to provide an estimate of the stacking energy.

In Section \ref{sec:model} we introduce the coarse-grained model of DNADs and we provide some details of the computer
simulations we performed. Section \ref{Sec:theory} gives a summary of the analytic theory which we developed in order to
describe the system in the isotropic and nematic phases. 
A comparison of our analytical approach with numerical results is presented in Section \ref{sec:resdisc}, while 
in Section \ref{sec:compwithexp} comparing our theoretical results with experimental data we provide an estimate
of the stacking energy. In Section \ref{sec:concl} we draw the conclusions of our work. 


\section{Model and Numerical Details}
\label{sec:model}
In this section we introduce a coarse-grained model devised to capture the essential physical 
features of end-to-end stacking (equilibrium polymerization) of DNA duplexes and well suited to being investigated both theoretically and numerically.
In the model, particles (DNADs) are assimilated to superquadrics (SQ) with a quasi-cylindrical 
shape decorated with two reactive sites on their bases 
determining their interactions.
SQs are a straightforward generalization of hard ellipsoids (HE), their surface is in fact defined as follows:
\begin{equation}
f(x,y,z) = \left |\frac{x}{a}\right |^p + \left |\frac{y}{b}\right|^m + \left |\frac{z}{c}\right |^n - 1 = 0
\label{Eq:SQdef}
\end{equation}
where the parameters $p,m,n$ are real numbers and $a$, $b$, $c$ are the SQ semi-axes. 
In our case we set $m=n=2$, $p=16$ and $b=c$, so that the SQ resembles a cylinder with rounded
edges (see Fig. \ref{Fig:singleSQ}).
Such SQs can be fully characterized by the elongation $X_0=a/b$ and by the parameter
$p$, that determines the sharpness of the edges (see Fig. \ref{Fig:singleSQ}).
SQs of elongation $X_0 < 1$ are called ``oblate'', while SQs of elongation $X_0 > 1$
are called ``prolate'', as for the HEs.
As unit of length in our simulations we use the length of short semi-axes $b$.
In the present study we investigated only prolate SQs with elongations $X_0=1,2$ and $3$.
We chose such elongations because DNADs used in experiments \cite{BelliniScience07} have a diameter $D=2 nm$ and are composed of $6$ to $20$ base pairs (BP) each $0.3$ nm long, hence their elongation $X_0$ ranges from $1$ to $3$.


Each particle is decorated with two attractive sites, located along the symmetry axis ($x$-axis in Fig. \ref{Fig:singleSQ}) at a distance $d/b =X_0 - 0.46$ from the DNAD center of mass, in order to model hydrophobic (stacking) forces between DNADs.
Sites belonging to distinct particles interact via the following square-well (SW) potential: 
\begin{equation}
\beta u_{\rm SW} \, = \,
\begin{cases}
-\beta \Delta E_S & r < \delta \\
0 & r > \delta
\end{cases}\:,
\label{eq:swpot}
\end{equation}
 where $r$ is the distance between the interacting sites, $\delta/b = 1.22 $ is the range of interaction (i.e. the diameter of the attractive sites), $\beta= 1/k_BT$ and $k_B$ is the Boltzmann constant. 
Therefore in the present model the anisotropic hard-core interaction is complemented with an anisotropic attractive potential in a fashion similar to what has been done in the past for other systems, like water \cite{simone}, silica \cite{DeMicheleSilica06} and stepwise polymerization of bifunctional diglycidyl ether of bisphenol A with pentafunctional diethylenetriamine \cite{CorezziSM08,corezziJPCB}.  
 
The location and diameter of the attractive sites have been chosen to best mimic the stacking interactions between blunt-ended DNAD 
and in particular they ensure that:
\begin{enumerate} 
\item the maximum interaction range
between two DNADs bases is of the order of typical range for hydrophobic interactions (i.e. $2 \AA$ see \cite{LeeHydropJCP84}), i.e. of the order of  water molecule dimensions 
\item the extent of the attractive surface of the DNADs bases is compatible with the surface of aromatic groups present in DNADs and which are responsible for hydrophobic interactions 
\end{enumerate}
We  note that in the present model each DNAD is symmetric around the $x-$axis (see Fig. \ref{Fig:singleSQ}), hence we are neglecting rotations around it.

We performed Monte Carlo (MC) simulations in the canonical and isobaric ensembles. We implemented the aggregation 
biased MC technique (AVBMC) developed by Chen and Siepmann \cite{SiepmannJPCB00,SiepmannJPCB01} in order
to speedup the formation of linear aggregates. 
To detect the overlap of two DNADs we calculated the distance  using the algorithm described in Ref. \cite{mioJCompPhys10}. 
In all simulations we adopted periodic boundary conditions in a cubic simulation box.
We studied a system of $N=1000$ particles in a wide range of volume
fractions $\phi$ and pressure $P$ respectively. 
Initially we prepared configurations at high temperature with all DNADs not bonded, then we quenched the system to the  final temperature (i.e. to the final value of $\beta\Delta E_S$) letting the system equilibrate. We checked equilibration by inspecting
the behavior of the potential energy and the nematic order parameter (see Sec. \ref{sec:nemphaseres}) in the system.

\section{Theory}
\label{Sec:theory}
Following the work of van der Schoot and Cates \cite{CatesLangmuir94,CatesEPL94} and its extension to higher volume fractions
with the use of Parsons-Lee approximation \cite{Parsons79,Lee87} as suggested by Kuriabova {\it et al.} \cite{GlaserMC},
we assume the following expression for the free energy of our system:
\begin{eqnarray}
\frac{\beta F}{V} &=& \sum_{l=1}^{\infty} \nu(l) \left \{ \ln \left [ v_d \nu(l) \right ] - 1 \right \} +\nonumber \\
&+& \frac{\eta(\phi)}{2} \sum_{l=1 \atop l'=1}^{\infty} \nu(l) \nu(l') v_{excl}(l, l')  
 \nonumber \\ 
 &-&  (\beta \Delta E_S  + \sigma_b) \sum_{l=1}^{\infty} (l-1) \nu(l)  
 + \sum_{l=1}^{\infty} \nu(l) \sigma_o(l) 
\label{eq:freeene}
\end{eqnarray}

where $\nu(l)$ is the number density of chain of length $l$, 
normalized such that $\sum_{l=1}^{\infty} l\, \nu(l)= \rho$, $v_d$ is the volume of a monomer,
$\beta \Delta E_S$ is the (positive) stacking energy, 
 $v_{excl}(l,l')$ is the excluded volume of two chains of length $l$ and $l'$ and $\sigma_b$ is the entropic free energy penalty for bonding (i.e.
is the contribution to free energy due to the entropy which is lost by forming a single bond) and $\eta(\phi)$ is the Parsons-Lee factor\cite{Parsons79}
\begin{equation}
\eta(\phi) = \frac{1}{4} \frac{4-3\phi}{(1-\phi)^2}
\label{eq:parsonlee}
\end{equation}
and $\sigma_o $\cite{Odijk86}  accounts for the orientational entropy that a chain of length $l$ loses in the nematic phase (including possible contribution due to its flexibility).
Differently from Ref.~\cite{CatesEPL94,GlaserMC}, but as in Ref.~\cite{KindtJCP04}, we  explicitly account for the polydispersity inherent in the equilibrium polymerization using a discrete chain length distribution. 
We explicitly separate the bonding free energy in an energetic ($\beta \Delta E_S$) and entropic
($\sigma_b$) contribution.  Differently from  Ref.~\cite{CatesEPL94,KindtJCP04}, but as in Ref.~\cite{GlaserMC} we include the Parson-Lee factor. Indeed, the Parson decoupling approximation satisfactory models  
the phase diagram of uniaxial hard ellipsoids \cite{AllenJCP96}, hard cylinders 
\cite{LekkMolPhys09}, linear fused hard spheres chains \cite{VargaMolPhys00}, mixtures of hard platelets \cite{LekkJPCB01}, hard sphero-cylinders \cite{LagoCPL10,CinacchiJCP04,SpheroCylRec},
rod-plate mixtures \cite{LekkJCP01}, mixtures of rod-like particles\cite{JacksonMolPhys03,JacksonJCP03} and mixtures of hard rods and hard spheres\cite{CuetosPRE07}. 
On the other hand,  Ref.~\cite{JacksonJCP98} finds that the Parsons theory is not satisfactory in the case of rigid linear chains of spheres. 

A justification of the use of Parsons-Lee factor in Eq. (\ref{eq:freeene}) 
for the present case of aggregating cylinders  is provided in Appendix~A.
Here we only note that the present system, in the limit of high $T$ where polymerization is not effective, reduces to 
a fluid of hard quasi-cylinders, where  the use of Parsons-Lee factor is justifiable\cite{LekkMolPhys09,CinacchiJCP04,SpheroCylRec}.  Moreover, in the dilute limit  ($\eta(\phi)\rightarrow 1$) 
the excluded volume term in Eq. \ref{eq:freeene}
reduces to the excluded volume of a polydisperse set of cluster with length distribution $\nu (l)$, which is conform to 
Onsager original theory \cite{Onsager49}.
In other words the form chosen in Eq. \ref{eq:freeene} for the excluded volume contribution to the free energy reduces to  the correct expressions in the limit of high temperatures and in that of low volume fractions.
  
Following Van der Schoot and Cates \cite{CatesLangmuir94,CatesEPL94} a generic form for $v_{excl}(l,l')$ can be assumed as a second order polynomial   in $l$ and $l'$, with 
\begin{eqnarray}
&&v_{excl}[l,l'; f({\bf u})] = 2  \int  f({\bf u}) f({\bf u}') D^3 \left [ \Psi_1(\gamma,X_0) + \right . \nonumber \\ 
&+&\frac{l + l'}{2} \Psi_2(\gamma,X_0) X_0 +\left. \Psi_3(\gamma,X_0) 
X_0^2\; l\, l' \right ] d{\bf\Omega}\, d{\bf\Omega}'
\label{eq:genvexcl}
\end{eqnarray}
where  $f({\bf u})$ is the probability for a given monomer of having orientation ${\bf u}$ within the solid ${\bf\Omega}$ and ${\bf\Omega}+d{\bf\Omega}$ and $\Psi_\alpha$ describe the angular dependence of the excluded volume.
The orientational probability $f({\bf u})$ are normalized as
\begin{eqnarray}
\int f({\bf u}) d\Omega = 1 
\label{eq:normcond}
\end{eqnarray}

In particular for two rigid chains of length $l$ and $l'$ which are composed of hard cylinders (HC) of diameter $D$ and length $X_0 D$ 
$v_{excl}(l,l')$  has been calculated by Onsager in 1949 
\begin{eqnarray}
v_{excl}(l,l') &=& \int f({\bf u}) f({\bf u}') D^3 \left [ \, \frac{\pi}{2} \sin\gamma + \frac{\pi}{2} X_0  ( 1 + |\cos\gamma|  + \right . \nonumber \\
&+& \left . \frac{4}{\pi} E(\sin\gamma) )  \frac{l  + l'}{2} + 2 X_0^2 \sin \gamma\;\; l\,l'\, \right ] d{\bf\Omega}\, d{\bf\Omega}'
\label{eq:vexclHC}
\end{eqnarray}
where $\cos\gamma={\bf u}\cdot{\bf u}'$ and $E(\sin\gamma)$ is the complete elliptical integral
\begin{equation}
E(\sin\gamma) = \frac{1}{4} \int_0^{2\pi} (1-\sin^2\gamma \sin^2\psi)^{1/2} d\psi  
\label{eq:ellipint}
\end{equation}
On passing we observe that the integrals in Eq. (\ref{eq:vexclHC}) can be calculated exactly in the isotropic phase 
while in the nematic phase the calculation can be done analytically only with suitable choices
of the angular distribution $f({\bf u})$.  
Comparing Eqs. (\ref{eq:vexclHC})  and (\ref{eq:genvexcl}) for HC one has:
\begin{eqnarray}
\Psi_1(\gamma,X_0) &=&  \frac{\pi}{2} \sin\gamma \nonumber\\
\Psi_2(\gamma,X_0) &=& \frac{\pi}{2} (1 + |\cos\gamma|  + \frac{4}{\pi} E(\sin\gamma) ) \nonumber\\
\Psi_3(\gamma,X_0) &=& 2 \sin\gamma
\label{eq:psiforHC}
\end{eqnarray}
In view of Eqs. (\ref{eq:psiforHC}) we notice that for HCs the functions $\Psi_1(\gamma)$, $\Psi_2(\gamma)$ and $\Psi_3(\gamma)$
accounts for the orientational dependence of the excluded volume of two monomers having orientations ${\bf u}$ and ${\bf u}'$
with ${\bf u}\cdot{\bf u}' =\cos\gamma$.
It is also worth observing that the first term of the integrand in Eq. (\ref{eq:vexclHC}) is independent of $l$ and hence accounts for the excluded volume interaction between two HCs end caps, the second term is linear in $l$ and $l'$ and accounts for the excluded volume between the two end caps of a chain and all midsections of the other one and the third one proportional to $l l'$ models the interaction  between all $l l'$ pairs of midsections 
of the two chains\cite{CatesLangmuir94,CatesEPL94}.
In summary,  Eq. (\ref{eq:genvexcl}) is exact for two rigid chains of HCs 
but according to \cite{CatesLangmuir94,Semenov81} to lowest order of approximation it is justifiable to use such equation also for two semi-flexible chains. We then assume that $v_{excl}$ remains additive with respect to end-end, end-midsection and midsection-midsection 
excluded volume contributions even if the chain is semi-flexible.
Finally our further ansatz is that Eq. (\ref{eq:genvexcl}) is also a good functional form for the excluded volume of two superquadrics having quasi-cylindrical shape: we will check the validity of this hypothesis using our simulations data. 

An exact expression for $\sigma_o$ is not available. The two following limits
have been calculated by Khokhlov and Semenov \cite{Semenov82,Odijk86,Vroege92}:
\begin{equation}
\begin{array}{llr}
\displaystyle\sigma_o(l) &\displaystyle= \frac{l}{8 l_p} \int \left ( \frac{df}{d\theta}\right )^2 f^{-1}d{\bf\Omega}\; +
\hspace{1cm} &\\[0.5cm]
& - 2 \ln\left [\int f^{1/2} d{\bf\Omega}\right ] + \ln(4\pi) & \hspace{1cm}(l_p \ll l)  \\[0.5cm]
\displaystyle \sigma_o(l) &\displaystyle= \int f \ln(4\pi f) d{\bf\Omega}\; + \hspace{1 cm} &\\[0.5cm]
& + \frac{l}{12 l_p} \int \left ( \frac{df}{d\theta}\right )^2 f^{-1}d{\bf\Omega}  & \hspace{1cm}(l_p \gg l)
\end{array}
\label{eq:semenov}
\end{equation}

Finally, we notice that  in the limit of rigid rods with $f_l({\bf u})=f({\bf u}) \nu(l)$, (the same limit 
selected in Ref.~\cite{GlaserMC}), the  free energy in Eq. (\ref{eq:freeene}) reduces to:
\begin{eqnarray}
\frac{\beta F}{V} &=& \frac{2}{3 l_p}\sum_{l=1}^{\infty} l  \int \left [f_l({\bf u})\right ]^{1/2} \nabla^2 
\left [f_l({\bf u})\right ]^{1/2} d{\bf\Omega} \nonumber \\
 &+&  \sum_{l=1}^{\infty} \int f_l({\bf u}) \left \{ \ln \left [ 4\pi  v_d f_l({\bf u}) \right ] - 1 \right \} +\nonumber \\
&+& \frac{\eta(\phi)}{2} \sum_{l=1,l'=1}^{\infty} \int f_l({\bf u}) f_{l'}({\bf u'}) v_{excl} d{\bf\Omega} d{\bf\Omega}' +  \nonumber\\
 &-& (\beta \Delta E_S  + \sigma_b) \sum_{l=1}^{\infty} \int (l-1) f_l({\bf u}) d{\bf\Omega} 
\label{eq:freeeneGlaser}
\end{eqnarray}
which is analogous to the free energy expression used by Glaser {\it et al.} \cite{GlaserMC}.


\subsection{Isotropic phase}
\noindent In the isotropic phase all orientations are equiprobable, hence:
\begin{equation}
f({\bf u}) = \frac{1}{4\pi}
\label{eq:isof}
\end{equation}
Plugging Eq. (\ref{eq:isof}) into Eq. (\ref{eq:freeene}) and calculating the integrals one obtains:
\begin{eqnarray}
\frac{\beta F}{V} &=&  \sum_{l=1}^{\infty} \nu(l) \left \{ \ln \left [v_d\nu(l)\right ] - 1 \right \} + \nonumber \\ 
&+& \frac{\eta(\phi)}{2} \sum_{l=1,l'=1}^{\infty} \nu(l) \nu(l') v_{excl}(l, l') + \nonumber\\
 &-& (\beta \Delta E_S  + \sigma_b) \sum_{l=1}^{\infty} (l-1) \nu(l)
\label{eq:freeene_iso}
\end{eqnarray} 

For hard cylinders the excluded volume can be calculated explicitly and it turns out to be:
\begin{eqnarray}
v_{excl}(l,l') &=& \frac{\pi^2}{8} D^3 + \left (\frac{3\pi}{8} + \frac{\pi^2}{8}\right) [l+l'] X_0 D^3 +\nonumber\\ 
&+& \frac{\pi}{2} l\,l' X_0^2 D^3 
\label{eq:vexclHCiso}
\end{eqnarray}

Building on Eq. (\ref{eq:vexclHCiso}), the generic expression for the excluded volume $v_{excl}(l,l')$
reported in Eq. (\ref{eq:genvexcl}) in the isotropic phase takes the form:
\begin{eqnarray}
v_{excl}(l,l') &=& 2 \left [ A_I(X_0) + k_I(X_0) v_d \frac{l+l'}{2} +\right .\nonumber\\
&& \left . \phantom{\frac{l+l'}{2}}B_I(X_0) X_0^2 l l' \right ] 
\label{eq:vexcl_iso}
\end{eqnarray}

We assume that the chain length distribution $\nu(l)$ is exponential with a average chain length $M$ 
\begin{equation}
\nu(l) = \rho M^{-(l+1)} (M-1)^{l-1} 
\label{eq:chaindistro}
\end{equation}
where
\begin{equation}
M = \frac{\sum_1^{\infty} l \, \nu(l)}{\sum_1^{\infty} \nu(l)}.
\label{eq:defM}
\end{equation}
With this choice for $\nu(l)$ the free energy in Eq. (\ref{eq:freeene_iso}) becomes:
\begin{eqnarray}
\frac{\beta F}{V} &=&  -\rho (\beta \Delta E_S + \sigma_b) (1 - M^{-1}) +\nonumber\\
&+& \eta(\phi) \left [ B_I X_0^2 + \frac{v_d k_I }{M} + \frac{A_I}{M^2}\right ]\rho^2  + \nonumber\\
&+&  \frac{\rho}{M} \left [ \ln\left ( \frac{v_d\rho}{M} \right )- 1\right ] + \nonumber\\
&+& \rho \frac{M-1}{M} \ln(M-1) - \rho \ln M.
\label{eq:Fiso}
\end{eqnarray}
Note that in general $k_I$, $B_I$ and $A_I$ depend on $X_0$.


The minimum of the  free energy 
with respect to $M$  (i.e. the equation ${\partial (\beta F / V) }/{\partial M} = 0 $ provides the searched equilibrium value for $M$. Dropping terms in 
$O(1/M^2)$ 
one obtains
\begin{equation}
M = \frac{1}{2} \left ( 1 + \sqrt{1 + 4 \omega \phi e^{k_I \phi \eta(\phi) + \beta \Delta E_S} }\right)
\label{eq:avgchain}
\end{equation}
where $\omega \equiv 4 e^{\sigma_b}$.
This formula for $M$  differs from the one reported by Kindt \cite{KindtJCP04}  by the
presence of the  Parsons-Lee factor, which will play a role at high volume fractions.
 
The expression for $M$ in Eq. (\ref{eq:avgchain}) coincides with  the parameter-free expression for average chain length $M_w$
obtained within Wertheim's theory (e.g. see Refs. \cite{WertheimJSP1,WertheimJSP2,WertheimJSP3,BianchiJCP07,JacksonMolPhys88}), when 
$\phi$ is small and $e^{k_I \phi \eta(\phi)} \approx 1$. Indeed, in Wertheim theory

\begin{equation}
M_{W} =  \frac{1}{2} + \frac{1}{2} \sqrt{1 + 8 \frac{\phi}{v_d} \Delta}
\label{eq:werth}
\end{equation} 
where  $\Delta = V_b (e^{\beta \Delta E_S} - 1)$ and 
$V_b$ is the bonding volume\cite{BianchiJCP07}. 
In the limit $e^{\beta \Delta E_S} \gg 1$, always valid in the $T$-region where chaining takes place,  \begin{equation}
M_W =   \frac{1}{2} + \frac{1}{2} \sqrt{1 + 8 \frac{V_b}{v_{d}}  \phi e^{\beta \Delta E_S}}
\end{equation}
The equivalence between the two expressions provide an exact definition of $\omega$ as 
\begin{equation}
\omega = 2 \frac{V_b}{v_{d}}
\label{eq:alpha_werth}
\end{equation}

Although  Eq. (\ref{eq:avgchain}) has been derived ignoring $O(1/M^2)$ terms in the free energy,
the average chain length $M$ can be always calculated, and this is what we do in this work, finding  numerically 
the zero of ${\partial (\beta F / V) }/{\partial M} = 0 $. 

\subsection{Nematic Phase}
In the nematic phase the function $f ({\bf u})$ depends explicitly on the angle between 
a given particle direction and the nematic axis, i.e. on the axis $\bf u$.
The function $f({\bf u})$ is also called the ``trial function'' and it 
generally depends on a set of parameters that have to be obtain through the minimization of the free
energy. 
Also in the nematic phase we assume an exponential distribution for $\nu(l)$.
In view of the analytical expression for the excluded volume $v_{excl}$ for cylinders, we assume the following form for the $v_{excl}$ of two DNADs averaged over the solid angle using a one parameter ($\alpha$) 
dependent trial function 
\begin{eqnarray}
v_{excl}(l,l',\alpha) &=& 2 \left [ A_N(\alpha) + v_d k_N(\alpha) \frac{l + l'}{2} +\right .\nonumber\\ 
&&\left .\phantom{\frac{l + l'}{2}}+ B_N(\alpha) X_0^2 l\,l' \;  \right ]   \hspace{0.5cm}
\label{eq:vexnem}
\end{eqnarray}


If we insert Eqs. (\ref{eq:xi}), (\ref{eq:vexnem}) and (\ref{eq:chaindistro})
into Eq. (\ref{eq:freeene}) we obtain after some algebra:

\begin{eqnarray}
\frac{\beta F}{V} &=&\hat \sigma_o - \rho (\beta \Delta E_S + \sigma_b) (1 - M^{-1}) +\nonumber\\ 
&+& \eta(\phi) \left [ B_N(\alpha) X_0^2 + \frac{k_N(\alpha)}{M} v_d + 
\frac{A_N(\alpha)}{M^2} \right ] \rho^2 +\nonumber\\
&+&  \frac{\rho}{M} \left ( \log\left[\frac{v_d\rho}{M}\right ]-1\right ) 
- \rho\log M + \nonumber\\
&+& \rho \log(M-1) \frac{M-1}{M}
\label{eq:Fnem}
\end{eqnarray} 
where $\hat \sigma_o \equiv \sum_l \sigma_o(l) \nu(l)$


\subsection{Phase Coexistence}
Using the free energy functionals in Eqs. (\ref{eq:Fiso}) and (\ref{eq:Fnem}) the phase boundaries, i.e. 
$\phi_N=v_d \rho_{N}$ and $\phi_I= v_d \rho_{I}$, of isotropic-nematic transition can straightforwardly calculated  by minimizing the free energy with respect to average chain lengths in the isotropic and nematic phases, i.e.
$M_{I}$ and $M_{N}$,  and $\alpha$. We also require  that  the isotropic and nematic 
phases have the same pressure, i.e. $P_{I}=P_{N}$ and the same chemical potential $\mu_{I}=\mu_{N}$.  
These conditions require numerically solving the following set of equations:
\begin{eqnarray}
\frac{\partial}{\partial M_{I}}  F_{iso} (\rho_{I}, M_{I}) &=& 0\nonumber\\
\frac{\partial}{\partial M_{N}}  F_{nem} (\rho_{N}, M_{N}, \alpha) &=& 0\nonumber\\
\frac{\partial}{\partial \alpha}  F_{iso} (\rho_{I}, M_{I},\alpha) &=& 0\nonumber\\
P_{I}(\rho_{I}, M_{I}) &=& P_{N}(\rho_{N},M_{N},\alpha)\nonumber\\
\mu_{I}(\rho_{I}, M_{I}) &=& \mu_{N}(\rho_{N},M_{N},\alpha)
\label{eq:phasecoexeqs}
\end{eqnarray} 

\section{Calculation of free energy parameters}
\label{sec:calcparam}
The theory illustrated in the previous section requires the calculation of several parameters, 
$V_b$, $k_I$, $A_I$, $B_I$,  $k_N$, $A_N$, $B_N$, $l_p$.
Since for super-quadrics an explicit calculation of these parameters is very unlikely in the following we describe simple methods to calculate them numerically.
For example the calculation of the excluded volume between clusters and the calculation of the bonding volume
require the evaluation of complicated integrals, which can be estimated with a Monte Carlo method.
The general idea indeed behind Monte Carlo is that such complicated integrals
can be calculated by generating a suitable distribution of points in the domain of integration.

\subsection{Excluded volume in the isotropic phase}
\label{sec:randinsertiso}

\begin{figure}[tbh]
\vskip 1cm
\includegraphics[width=.495\textwidth]{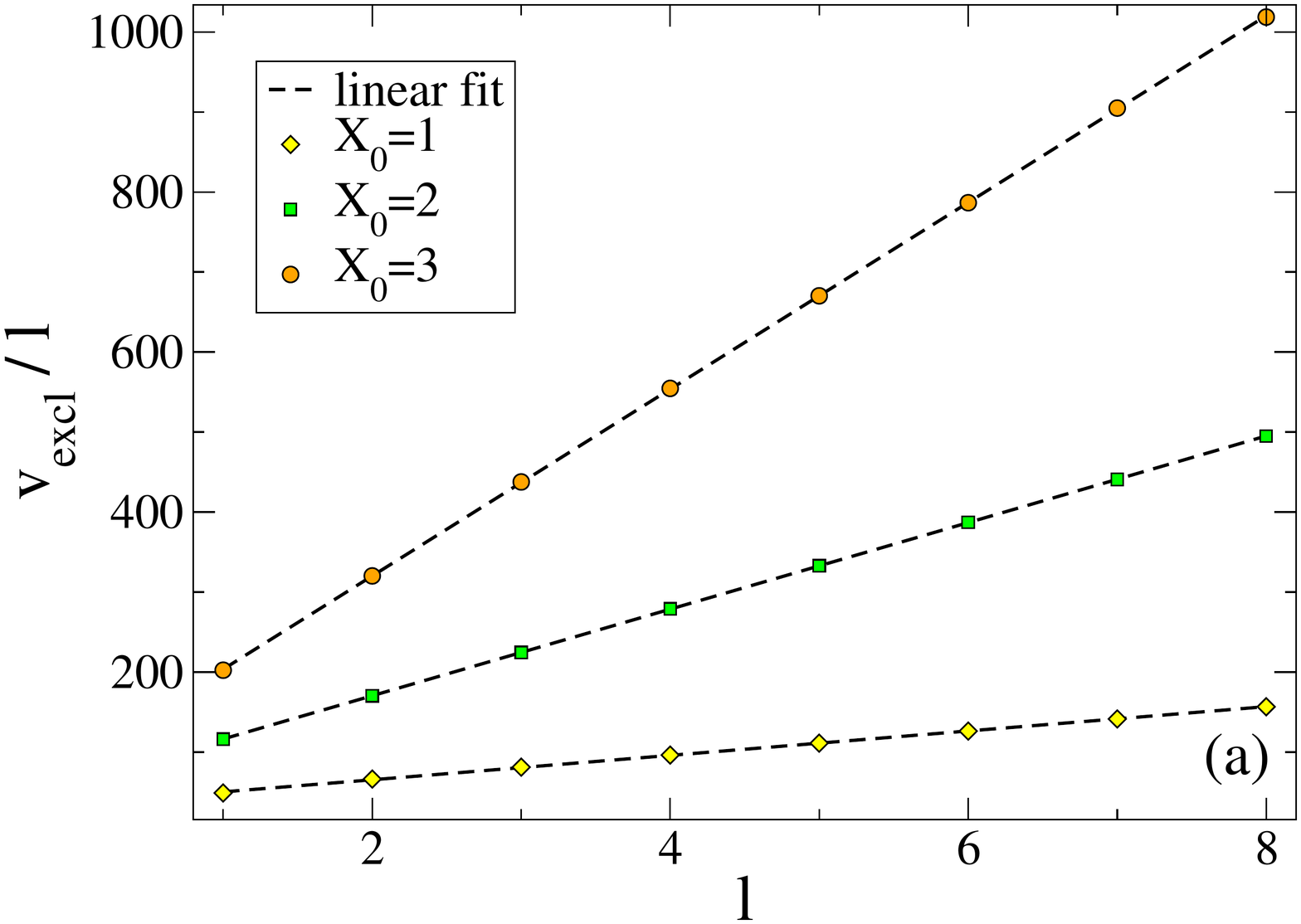}
\includegraphics[width=.495\textwidth]{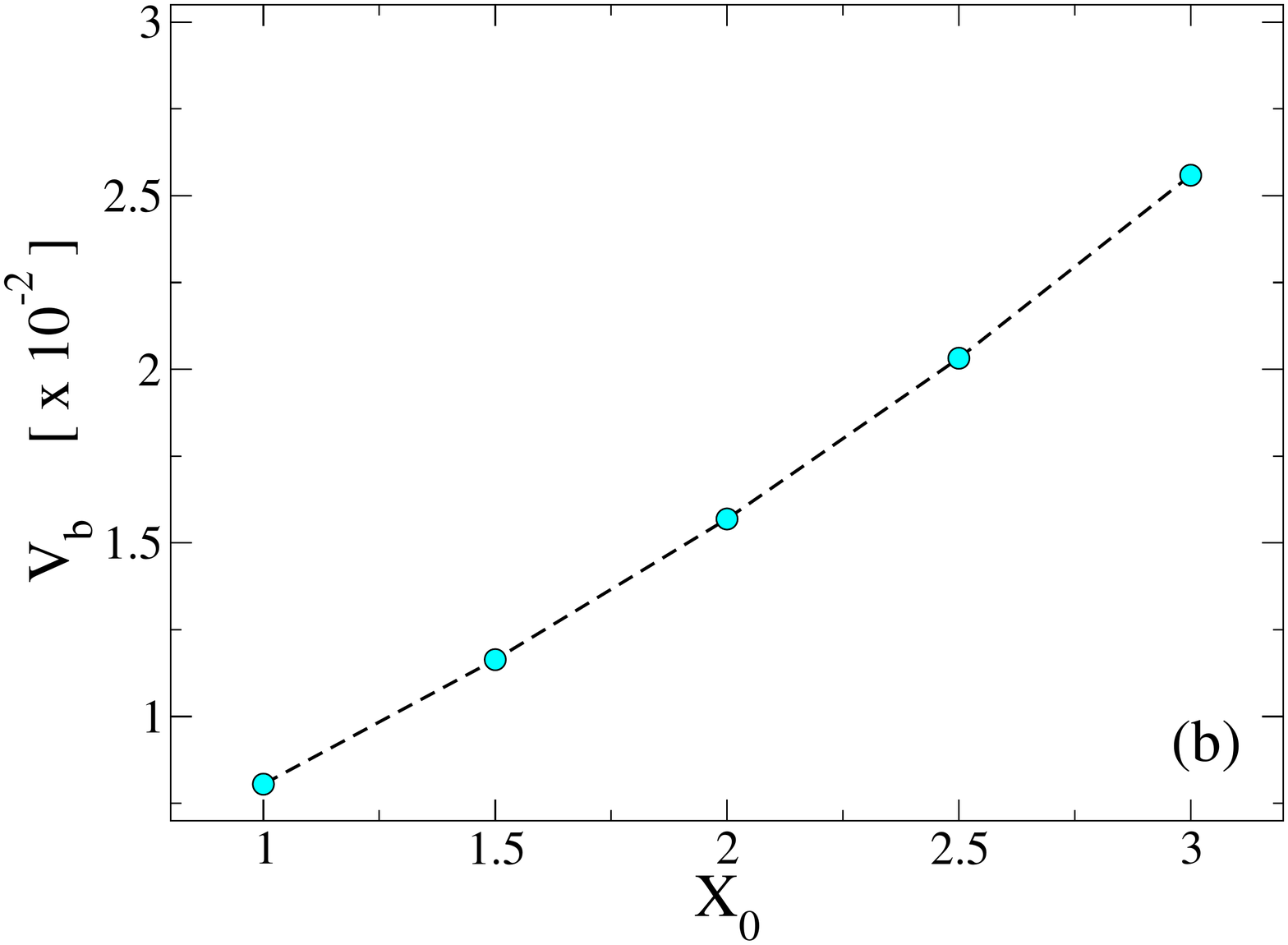}
\caption{(a) Excluded volume of two chains of  length $l$ calculated numerically as a function of $l$ for $X_0=1,2,3$.  Dashed lines are fits to Eq. (\ref{eq:vexSQ}). (b) Bonding volume as a function of elongation $X_0$.}
\label{Fig:covandvbond}
\end{figure}

In the isotropic phase, $v_{excl}(l,l')$ can be written as reported in   
 Eq. (\ref{eq:vexcl_iso}). If $l=l'$, 
\begin{equation}
v_{excl}(l,l) = 2 A_I  + 2 k_I v_d \, l + 2 B_I X_0^2  \,l^2
\label{eq:vexSQ}
\end{equation}
Hence, from a numerical evaluation of $v_{excl}(l,l)$ for several $l$ values (whose 
detailed procedure is described in Appendix \ref{AppendixB}) it is possible to estimate $A_I$, $k_I$ and $B_I$.
Fig. \ref{Fig:covandvbond} shows $v_{excl}(l,l)/l$  vs $l$. A straight line describe properly the data for
all $X_0$ values, 
suggesting that $A_I \approx 0$.  From a linear fit one obtain $2 B_I X_0^2 $  (slope) and 
$2 k_I v_d$ (intercept). 


\subsection{Calculation of the Bonding Volume}
The bonding volume $V_b$ can be calculated numerically performing a
Monte Carlo calculation of

\begin{eqnarray}
V_b &=& \int \theta\left (-\Delta E_S - u_{SW} - V_{HC}\right ) \,d{\bf r}\, d{\bf \Omega}_1\, {\bf\Omega}_2
\label{eq:vbondingw} 
\end{eqnarray} 
where $V_{HC}=V_{HC}({\bf r}, {\bf \Omega}_1,{\bf\Omega}_2)$ is the hard core part of the interaction potential and $\theta(x)$ is the Heaviside step function, i.e. $\theta(x) = 1$ if $x \ge 0$ or $0$ otherwise.
The detail of the numerical integration are reported in Appendix \ref{AppendixB}. 
The resulting values of $V_b$ for different $X_0$  are shown in 
Fig. \ref{Fig:covandvbond}(b). $V_b$ grows with $X_0$, an effect introduced by the different
rounding of the SQ surface close to the bases.    Indeed, how is it shown in Fig.~\ref{Fig:singleSQ},
on increasing $X_0$ the base surface is more rounded and such 
different rounding offers a different 
angular width over which bonds can form, an effect which will also reflect in the  $X_0$
dependence of the the persistence length of the self-assembled chains, as it will be discussed in details in Sec.~\ref{sec:perslen}.  The dependence of the bonding angle  on $X_0$ is not present for HCs
and in that case the bonding volume would be constant.

\subsection{Nematic phase}

In what follows we assume that the angular distribution of  the particle orientation can be well
described by the one parameter Onsager trial function\cite{Onsager49}, i.e.:
\begin{equation}
f({\bf u}) = f_O({\bf u}) = \frac{\alpha}{4 \pi \sinh\alpha} \cosh(\alpha \cos\theta)
\label{eq:fons}
\end{equation}
where $\theta$ is the angle between the particle and the nematic axis and the system is supposed to have azimuthal symmetry around such axis. 
The excluded volume $v_{excl}(l,l,\alpha)$ between two clusters of equal length $l$ can be calculated
using the procedure illustrated previously for the isotropic case with the only difference that now
monomers are inserted with an orientation extracted from the Onsager angular distribution defined in Eq.  (\ref{eq:fons}).

To estimate numerically $ A_N(\alpha)$, $k_N(\alpha)$ and $B_N(\alpha)$ we specialize Eq.~(\ref{eq:vexnem})
to the case of $l=l'=2$, $l=l'=3$, $l=l'=4$, evaluating numerically for several values of $\alpha$ 
$v_{excl}(2,2,\alpha)$, $v_{excl}(3,3,\alpha)$ and $v_{excl}(4,4,\alpha)$.  Inverting Eq.~(\ref{eq:vexnem})
allows us to express $ A_N(\alpha)$, $k_N(\alpha)$ and $B_N(\alpha)$  as a function of
$v_{excl}(2,2,\alpha)$, $v_{excl}(3,3,\alpha)$ and $v_{excl}(4,4,\alpha)$ as explained in detail in
Appendix~C.

\subsection{Estimate of the orientational entropy in the nematic phase}
We propose to model the  orientational entropy in the nematic phase using the following expression
proposed by Odijk \cite{Odijk86} (other possibilities can be found in Refs. \cite{DupreJCP91} and \cite{HentschkeMacromol90})

\begin{eqnarray}
\hat\sigma_o^{od} &=& \sum_{l=1}^{l=\infty} \nu(l) \left \{ \ln\alpha +\frac{(\alpha-1)\, l}{6 \, l_p}\right .\nonumber\\
 &+& \left .\frac{5}{12} \ln\left [\cosh\left (\frac{(\alpha-1)\, l}{ 5\, l_p}\right )\right ] - \frac{19}{12} \ln 2 \right \}
\label{eq:orientodijk} 
\end{eqnarray}


This expression, in the limit of  ``rigid chains''  (RC) and  ``flexible chains'' (FC)  reduces approximately to
the exact limits~\cite{Semenov81}

\begin{equation}
\begin{array}{llr}
\displaystyle\sigma_o^{RC}(l) &\displaystyle= \log(\alpha) - 1+  \frac{\alpha - 1}{6 l_p} l  &\hspace{1cm} \alpha l \ll l_p \\
\displaystyle\sigma_o^{FC}(l) &\displaystyle= \log(\alpha/4) + \frac{\alpha -1}{4 l_p} l  &\hspace{1cm} \alpha l \gg l_p \\
\end{array}
\label{eq:oriententnem}
\end{equation}

Latter formulas can be also obtained plugging the Onsager trial function $f_O({\bf u})$ in Eqs. (\ref{eq:semenov}).

Unfortunately,  Eq.~(\ref{eq:orientodijk})  is hardly tractable in the minimization procedure requested to
evaluated the equilibrium free energy and hence the two approximations in Eq.~(\ref{eq:oriententnem}) are
often preferred.  While in the case of fixed length polymers, the
knowledge of the persistence length selects one of the two expressions, in the case of 
equilibrium polymers, different chain length will contribute differently to the orientational entropy.
In particular, when the chain length distribution is rather wide, it is difficult to assess if the $RC$ 
(chosen in Ref.~\cite{GlaserMC})
or the $FC$ (chosen in Ref.~\cite{KindtJCP04})
limits should be used. 
 To overcome the numerical problem, still retaining both the $RC$ and the $FC$ behaviors, we  use the following expression for $\hat\sigma_o$:
\begin{eqnarray}
\hat\sigma_o &=& \sum_{l=1}^{l=l_0-1}  \nu(l) \left \{ \left [\log(\alpha) - 1\right ] +\phantom{\frac{\alpha - 1}{6 l_p}} \right .\nonumber\\
&+& \left .\frac{\alpha - 1}{6 l_p} \, l \right\} + \sum_{l=l_0}^{l=\infty} \nu(l)\left\{ 
 \log(\alpha/4) + \frac{\alpha -1}{4 l_p} \, l
\right\}
\label{eq:ourorientent}
\end{eqnarray}
in which the contribution of chains of size $l_0$ is treated with the $RC$ while the contribution of
longer chains enters with the $FC$ expression.  We pick $l_0$  by requesting maximum likeihood  
between Eq.~(\ref{eq:ourorientent}) and Eq.~(\ref{eq:orientodijk}) in the relevant $M$-$\alpha$ domain.  We found that the value $l_0 \approx 9$ is appropriate for most studied cases.


\subsection{Estimate of persistence length}
\label{sec:perslen}
In order to estimate the persistence length, entering in Eq.~(\ref{eq:ourorientent}),  we randomly  build chains according to the procedure
described in Appendix~B.
We estimate the ``chain persistence length'' $l_p$ by evaluating following spatial correlation function:
\begin{equation}
C_O(\left | i-j\right |) \equiv \sum_{i,j} \langle \hat{\bold x}(i)\cdot \hat{\bold x}(j) \rangle  
\label{Eq:chainpersist}
\end{equation}
where $i,j$ label two monomers along the chain ($i=0$ is the first monomer at chain end) and
$\hat{\bold x}(i)$ is a unit versor directed along $x$-axis of the monomer (i.e. their axis of symmetry, see Fig. \ref{Fig:singleSQ}), that coincides with the direction along which the two attractive sites lie.
$\langle \ldots \rangle$ denotes an average over the whole set of independent chains which has been generated.

\begin{figure}[tbh]
\vskip 1cm
\includegraphics[width=.49\textwidth]{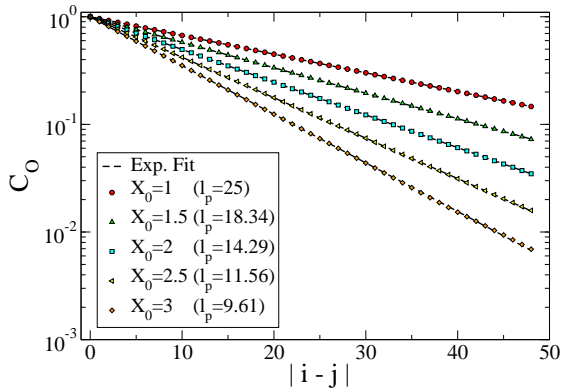}
\caption{Spatial correlation function $C_O(\left |i-j\right |)$ (see text for its definition) calculated generating random chains of $50$ monomers for aspect ratios $X_0=1,1.5,2,2.5,3$. Dashed lines are fits to the functional form $C_O(\left | i-j\right |) = \exp[-\left | i-j\right | / l_p]$. From these fits the chain persistence length $l_p$ can be estimated (see legend).  
}\label{Fig:persistence}
\end{figure}

In Figure \ref{Fig:persistence} we plot $C_O(\left | i-j\right |)$ for all elongations studied. All correlations decay
following an exponential law, whose characteristic scale is identified as the persistence length (in unit of monomer).
In the explored $X_0$ range, $10 < l_p < 25$.  The more elongated monomers have a smaller persistence length.
The  $X_0$ dependence of $l_p$ arises from the different roundness of the bases (implicit in the use of SQ),
as discussed in the context of the bonding volume and in Fig. \ref{Fig:singleSQ}.

\section{Results and Discussion}
\label{sec:resdisc}
In this section we compare results from simulations with theoretical calculations based
on the theory discussed in Sec. \ref{Sec:theory}

\subsection{Isotropic phase}
 
\begin{figure}[tbh]
\vskip 1cm
\begin{center}
\includegraphics[width=.49\textwidth]{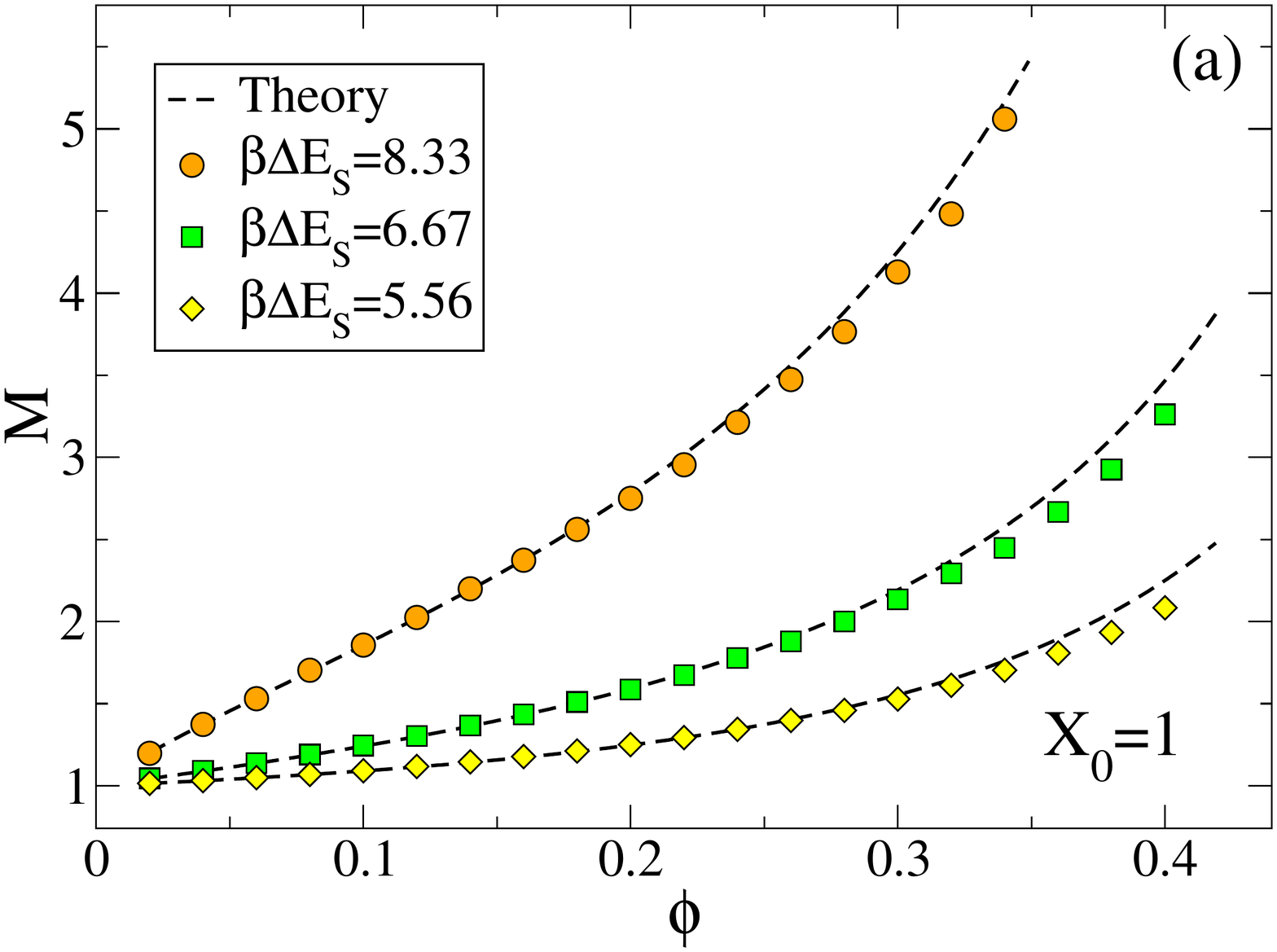}
\includegraphics[width=.49\textwidth]{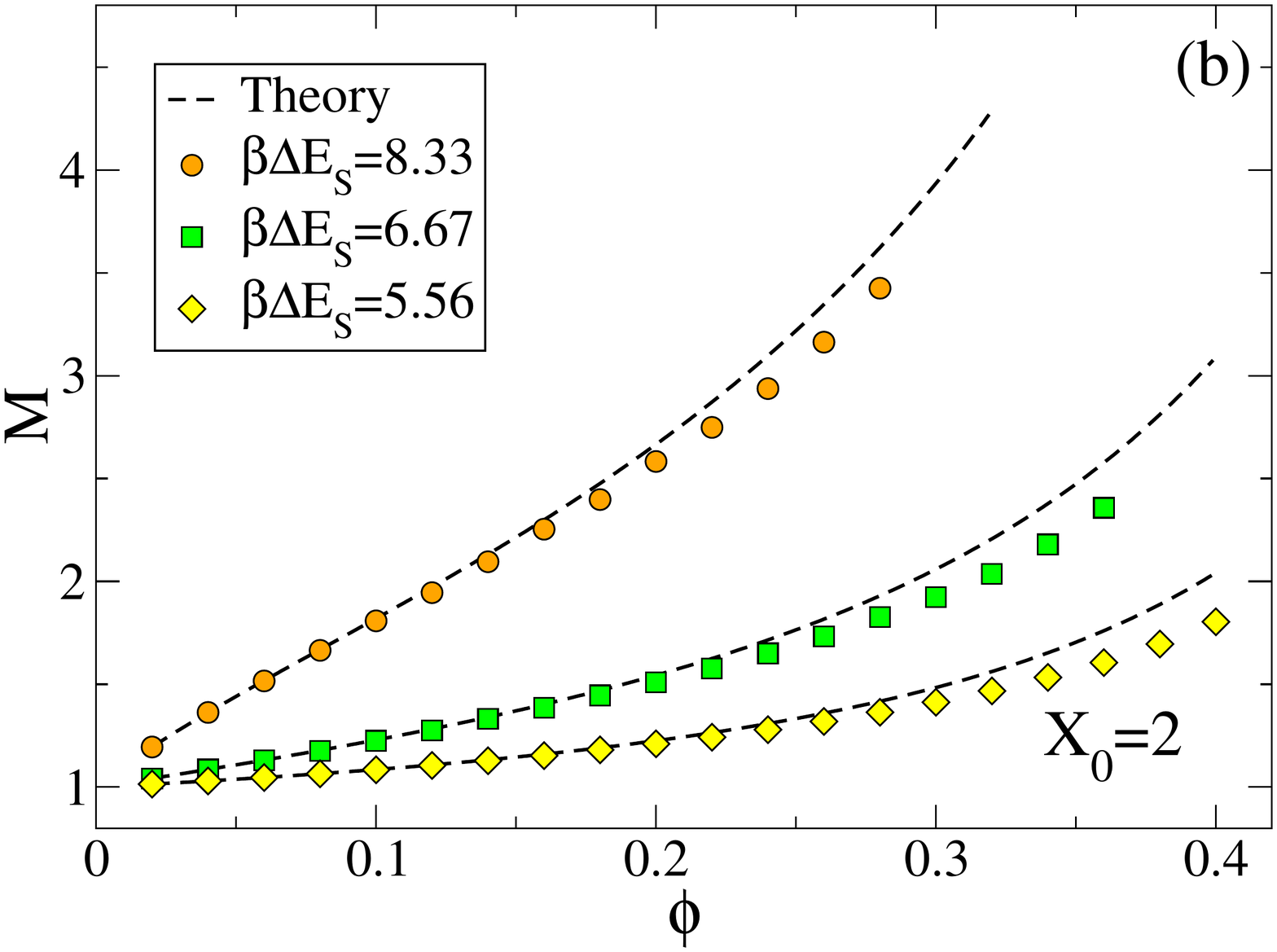}
\includegraphics[width=.49\textwidth]{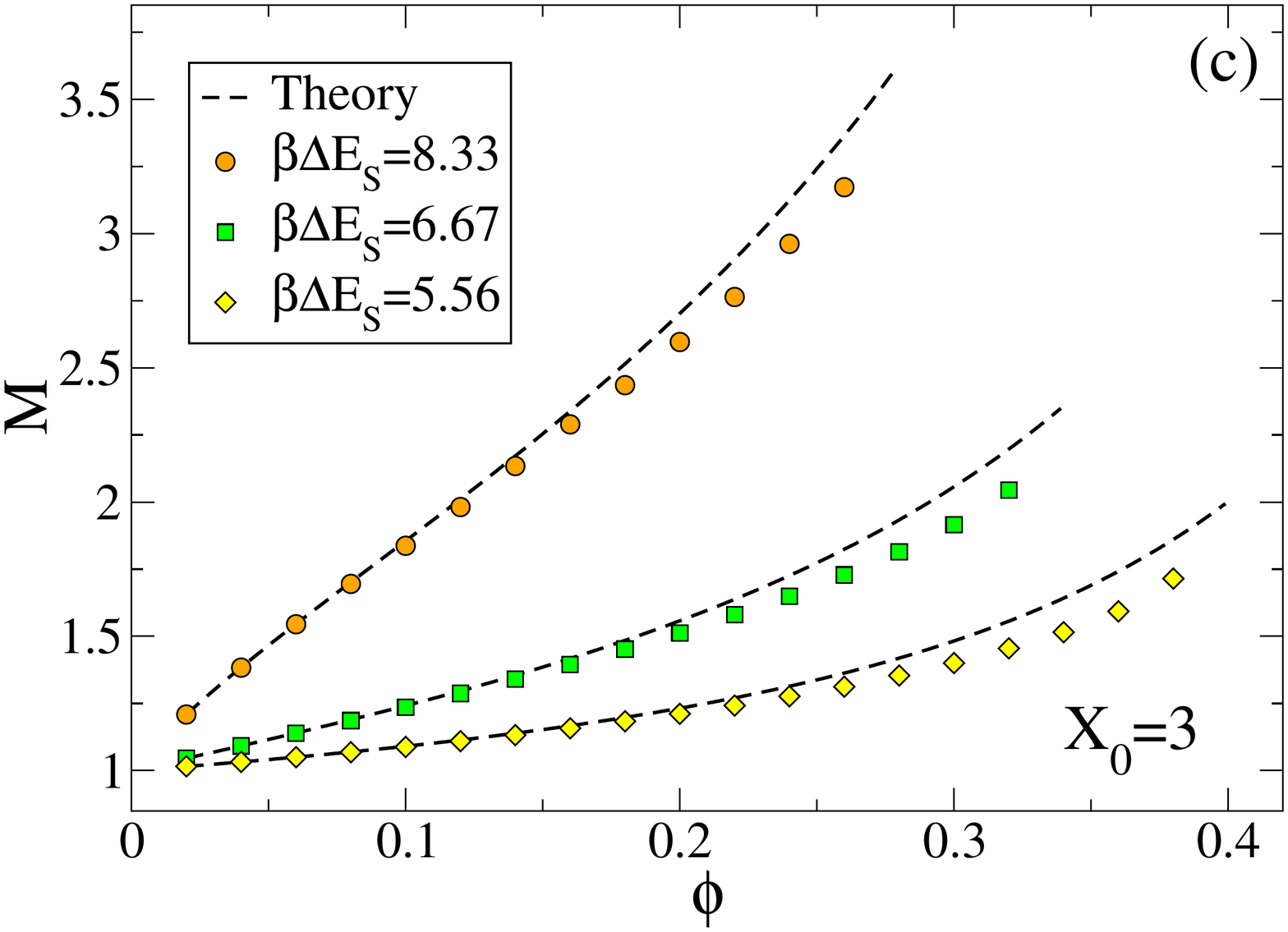}
\includegraphics[width=.49\textwidth]{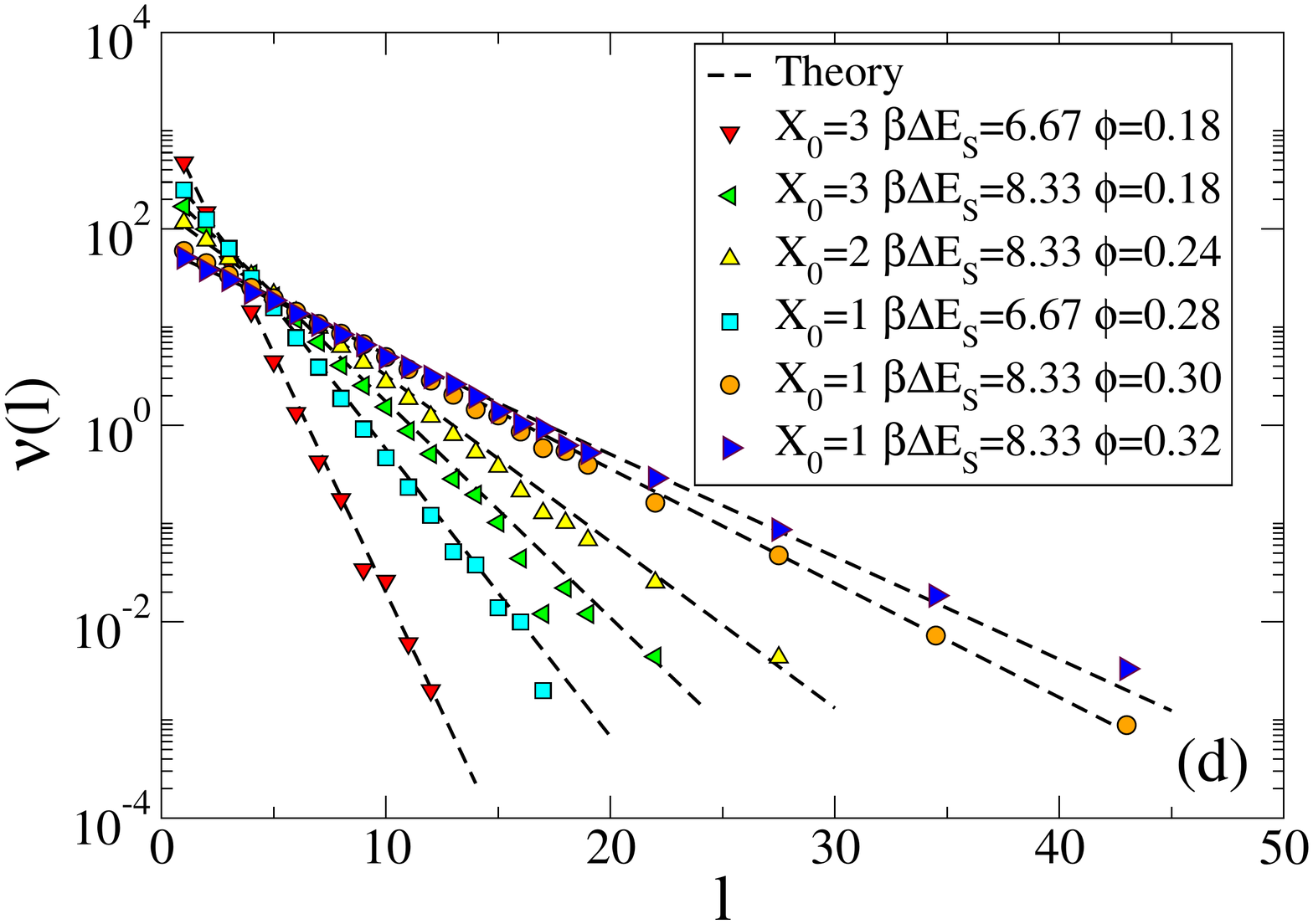}
\caption{(a)-(c) Average chain length $M$ against $\phi$ for $X_0=1,2,3$ at
all values of $\beta \Delta E_S$ studied. Dashed lines are theoretical predictions calculated by minimizing the free energy
in Eq. (\ref{eq:Fiso}) with values of $V_{b}$, $k_I$ and $B_I$ derived by the procedure described in Sec. \ref{sec:calcparam}.
(d) Cluster size distributions (colored symbols) for several state points together with theoretical predictions (dashed lines).}
\label{fig:MforAll}
\end{center}
\end{figure}

Fig. \ref{fig:MforAll} (a)-(c) show the packing fraction dependence of  $M$  for $X_0=1,2,3$ for all temperatures investigated. 
The dashed curves are calculated by minimizing with respect to $M$ the isotropic free energy in Eq.  (\ref{eq:Fiso}) using the values of $V_b$, $k_I$ and $B_I$ obtained in Sec. \ref{sec:randinsertiso} without any fitting parameter.
Up to volume fractions around $\phi\approx0.20$ the agreement between theoretical and numerical results
is quite good for all cases considered. Above such volume fraction the theoretical predictions start
deviating appreciably, a discrepancy that we attribute at
moderate and high $\phi$ to the inaccuracy of the Parsons decoupling approximation.   

In Fig. \ref{fig:MforAll} (d) we report the cluster size distribution $\nu(l)$ as obtained from both simulation and  
theory, the latter calculated according to Eq. (\ref{eq:Fiso}) with $M$ obtained by minimization of the isotropic free energy. 
As expected, the cluster size  
in  the isotropic phase is  exponential.     These results suggest that a reasonable first principle
description of the isotropic phase is provided by the free energy of Eq.~(\ref{eq:Fiso}), when the
parameter of the model are properly evaluated.  

\subsection{Nematic phase}
\label{sec:nemphaseres}
\begin{figure}[tbh]
\vskip 1cm
\begin{center}


\includegraphics[width=.45\textwidth]{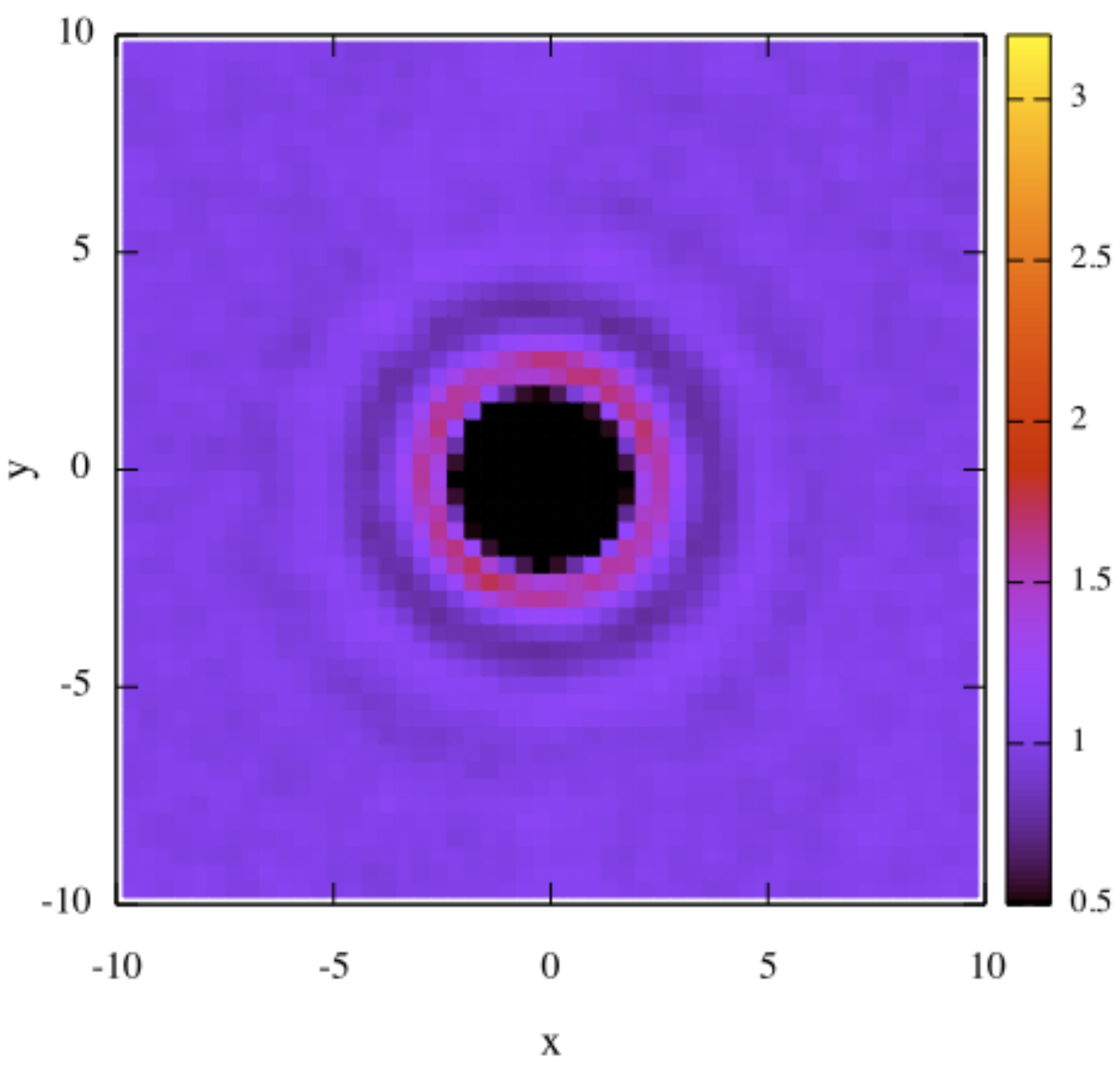}\hskip -0.5cm (a)
\includegraphics[width=.45\textwidth]{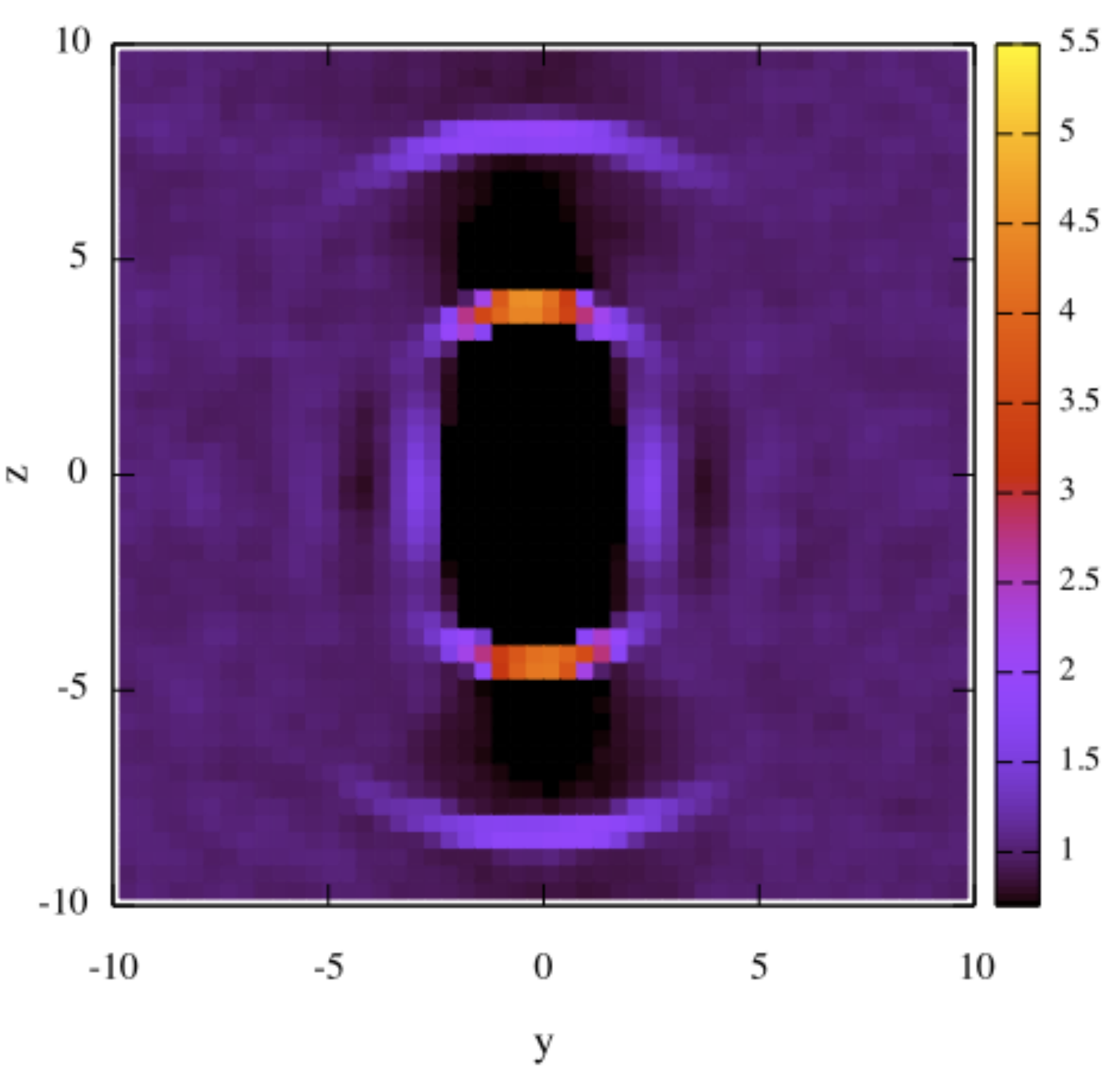}\hskip -0.5cm (b)
\hskip 0.5cm
\includegraphics[width=.45\textwidth]{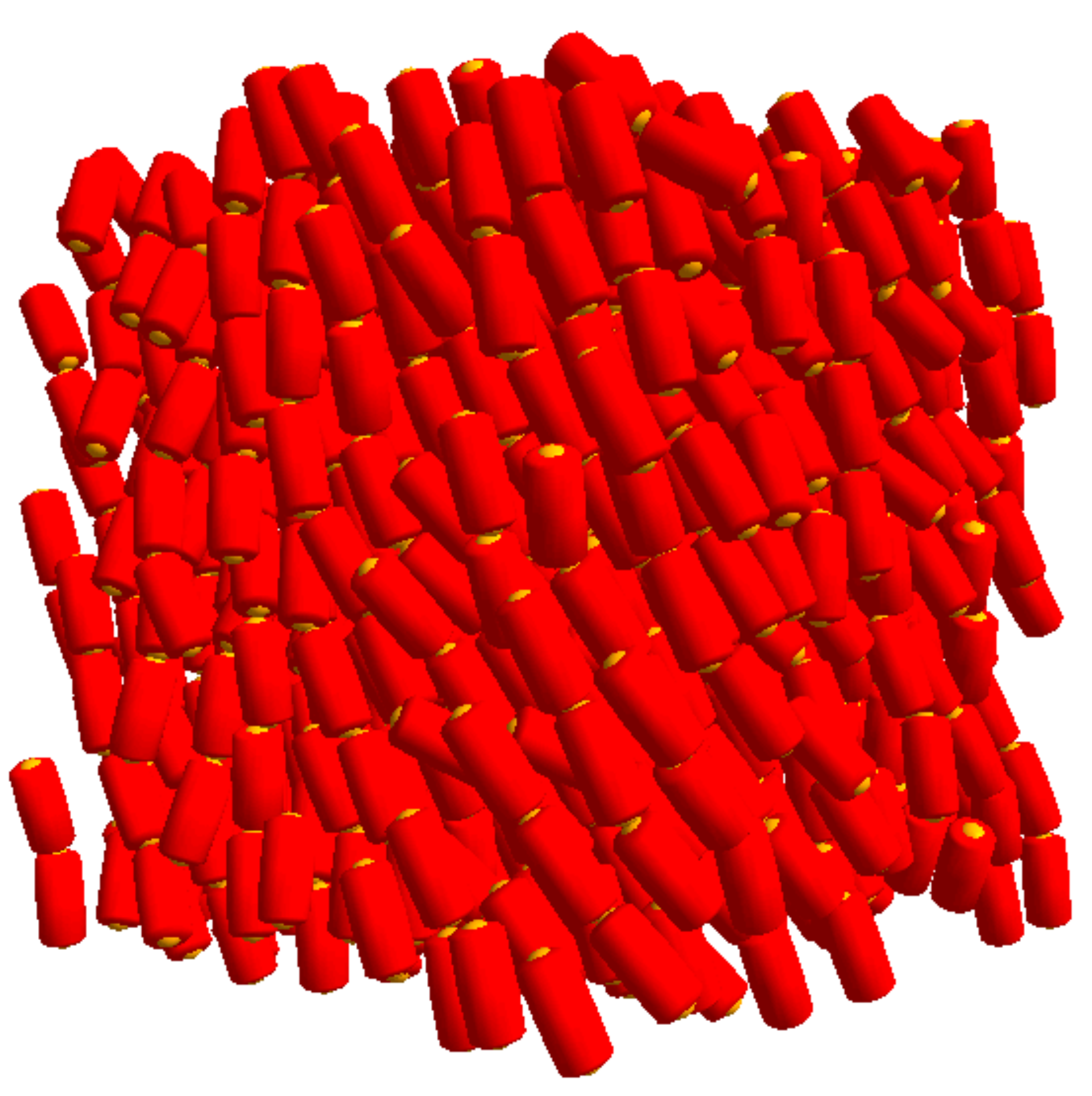}\hskip -0.5cm (c)
\caption{Plot of $g(x,y,0)$ (a) and $g(0,y,z)$ (b) where the $z$-axis is chosen parallel to the nematic director
for $X_0=2$, $\phi=0.38$ and $\beta \Delta E_S=8.33$.  (c) Example of nematic configurations at the same state point.}
\label{fig:rdf3d}
\end{center}
\end{figure}

On increasing $\phi$ the system transform into a LC phase. We estimate the degree of nematic ordering by evaluating the largest eigenvalue $S$ of the order tensor ${\bf Q}$, whose components are:
\begin{equation}
Q_{\alpha\beta} =\frac{1}{N}\sum_i  \frac{3}{2}\langle({\bf u}_i)_{\alpha} ({\bf u}_i)_{\beta}\rangle - \frac{1}{2} \delta_{\alpha,\beta}
\end{equation}
where $\alpha\beta\in\{x,y,z\}$, and the unit vector $({\bf u}_i(t))_\alpha$ is the component $\alpha$
of the orientation (i.e. the symmetry axis) of particle $i$ at time $t$.  A non-zero value of $S$
signals the presence of orientational order in the system and it can be found not only 
in the nematic  phase but also in partially ordered phases as columnar and smectic phases. Since in this article we focus
only on the nematic phase, to verify that the simulated state points are not partially ordered 
we calculate, following Ref. \cite{GlaserMC}, the three dimensional pair distribution function $g({\bf r})$ 
defined as:
\begin{equation}
g({\bf r}) = \frac{1}{\rho N} \left \langle  \sum_{i=1}^{N} \sum_{j\ne i} \delta({\bf r} - 
({\bf r}_i -{\bf r}_j))\right\rangle
\label{eq:rdf3d}
\end{equation}
where $\delta({\bf x})$ is the Dirac delta function.
We calculate the $g({\bf r})$ in a reference system with the $z$-axis parallel to the nematic
director. 
Figure \ref{fig:rdf3d}  shows  $g(x,y,0) $ and $g(0,y,z)$,  which  correspond respectively to 
the correlations in a plane perpendicular to the nematic director and in a plane containing it for a given nematic state point  ( $X_0=2$, $\phi=0.38$, $\beta \Delta E_S=8.33)$).   The  $g(x,y,0) $ is found to be isotropic, ruling out the possibility of a columnar or crystal phase (no hexagonal  symmetry is indeed present).
The $g(0,y,z)$ reflects the orientational ordering along the nematic direction and rules out
the possibility of a smectic phase (no aligned sequence of peaks are present\cite{GlaserMC}).
Fig.~\ref{fig:rdf3d}(c) shows also a snapshot of the simulated system at the same state point. 

In what follows, we have systematically calculated and inspected $g({\bf r})$ to verify that all state points having a value of $S$ large enough to be considered nematic are indeed translationally isotropic, i.e. with no translational order.

\begin{figure}[tbh]
\vskip 1cm
\begin{center}
\includegraphics[width=.45\textwidth]{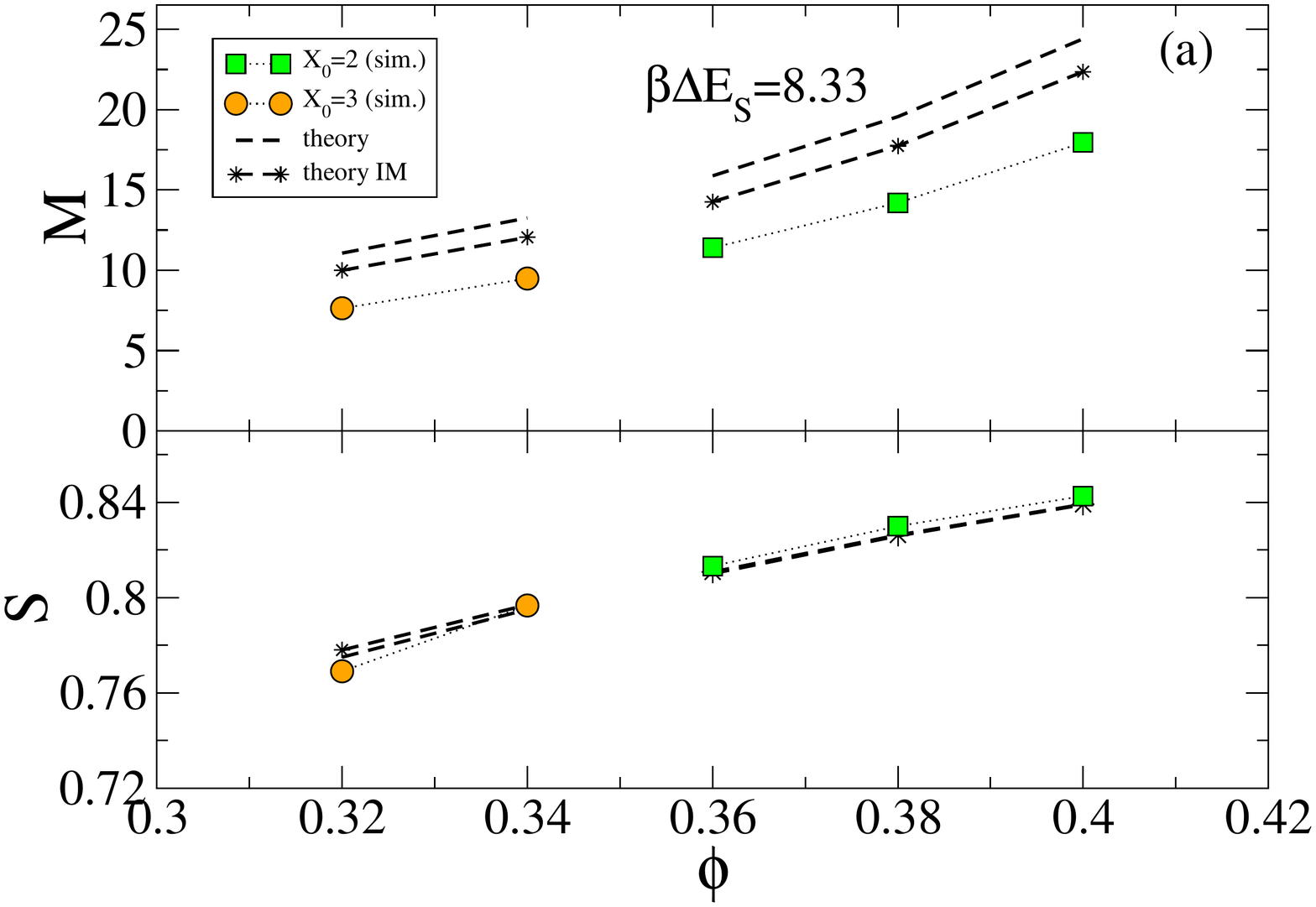}
\includegraphics[width=.45\textwidth]{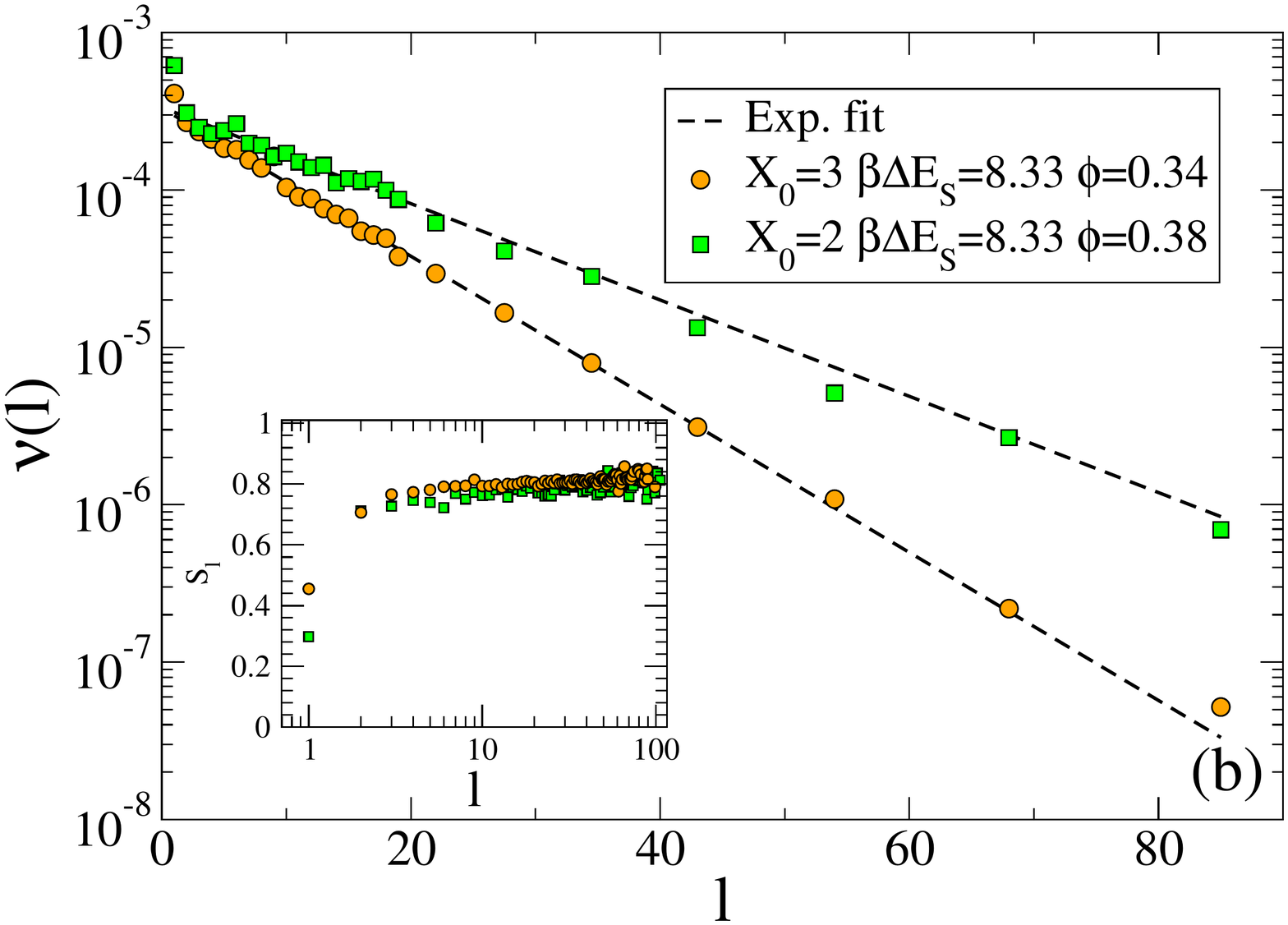}
\caption{(a) Average chain length and nematic order parameter $S$ for several nematic state points. Dashed lines with stars (theory IM) are theoretical predictions assuming that monomers are isotropic (see text for details).
(b) Cluster size distribution for two state points $(X_0=2,\phi=0.38,\beta \Delta E_S = 8.33)$ and $(X_0=3,\phi=0.34,\beta \Delta E_S = 8.33)$. Circles are numerical results and dashed lines are exponential fits. The inset shows  the chain length dependent nematic order parameter $S_l$ for the same state points.}
\label{fig:clsdstnem}
\end{center}
\end{figure}

 Fig. \ref{fig:clsdstnem} shows the nematic order parameter and  the average chain length $M$  calculated from simulations as well as  with the theoretical methodology described previously for two different elongations at
 $\beta \Delta E_S=8.33$.  
 %
 The theoretical value for $S$ is obtained according to:
 \begin{equation}
S(\alpha) = \int 2 \pi \frac{3 \,\cos^2\theta-1}{2} f_{O}(\theta; \alpha ) \sin\theta \;d\theta 
 \end{equation}  

 Fig. \ref{fig:clsdstnem} (a) shows that the nematic order parameter is very well captured by the theory,
 while the average chain length shows a clear disagreement between theory and simulations, again suggesting
 that the error introduced by the Parsons decoupling approximation which was already observed in the isotropic case for large packing fractions is here enhanced by the  further increase in $\phi$. 
Another possible source of error could arise from the hypothesis that the
cluster size distribution is exponential  also in the nematic phase. To test this hypothesis
we show in Fig.~\ref{fig:clsdstnem} (b) the cluster size distributions
in two different state points.   In all cases, the distributions are not a single exponential.  This phenomenon has been already observed and discussed by Lu and Kindt \cite{KindtJCP04} and described as  a two exponential decay of $\nu(l)$ with the exponential decay of short chains extending up to $l\approx 50$. They took into account  such bi-exponential nature of the distribution  to better reproduce the isotropic-nematic phase boundaries \cite{KindtJCP06} in their theoretical approach.  In the present case, only very short chains (not to say only the monomers),
fall out of the single exponential decay.  To test if the different decay reflects a different orientational ordering of the
small clusters compared to long chains, we follow Ref.~\cite{KindtJCP06} and  evaluate the length-dependent nematic order parameter $S_l$, that is  the nematic order parameter calculated for each population of clusters of size $l$. The results, reported in the inset of Fig. \ref{fig:clsdstnem}(b), show that 
$S_l$  is around $0.7-0.8$ for all clusters sizes except for $l=1$, i.e. except for monomers.  

To assess how much the theoretical predictions are affected by the assumption of a single exponential
decay (and of the associated identity of $S$ for all chains), we evaluate the  
 correction of the free energy functional in Eq. (\ref{eq:Fnem})  arising from the assumption that
 monomers are isotropic, while all other chains are nematic. To do so we 
exclude from the calculation of the orientational entropy the monomers and take into account
in the calculation of the excluded volume contribution the fact that monomers are isotropic. 
The revised free energy can be thus written as 
\begin{eqnarray}
\frac{\beta F}{V} &=& \hat\sigma_o^* - \rho (\beta \Delta E_S + \sigma_b) (1 - M^{-1}) + \nonumber\\
 &+&  \eta(\phi) \left [ B_N(\alpha) X_0^2 + \frac{k_N(\alpha)}{M} v_d + 
\frac{A_N(\alpha)}{M^2} \right ] \rho^2 +\nonumber\\
&-& \beta\Delta f_{N} + \frac{\rho}{M} \left ( \log\left[\frac{v_d\rho}{M}\right ]-1\right ) 
- \rho\log M + \nonumber\\
&+&\rho \log(M-1) \frac{M-1}{M}
\nonumber\\
\label{eq:correctedFnem}
\end{eqnarray} 
where 
\begin{eqnarray}
\sigma_o^* &=&  \sum_{l=2}^{l=l_0-1}  \nu(l) \left \{ \left [\log(\alpha) - 1\right ]+ \frac{\alpha - 1}{6 l_p} \, l \right\} +\nonumber\\
&+& \sum_{l=l_0}^{l=\infty} \nu(l)\left\{ \log(\alpha/4) + \frac{\alpha -1}{4 l_p} \, l
\right\}
\end{eqnarray}
and 
\begin{eqnarray}
\beta \Delta f_{N} &=& \eta(\phi) \left\{ v_d [ k_N(\alpha) - k_I ] \left (\frac{1}{M^2}+\frac{1}{M^3}  \right ) +\right .\nonumber\\
&+&\left . 2 X_0^2 [  B_N(\alpha) - B_I ]  \frac{1}{M^2} + 2 \frac{A_N(\alpha)}{M^3} \right\} \rho^2
\end{eqnarray}

Minimizing such expression we calculate the resulting improved estimate for the  average chain length and nematic
order parameter and the results are also shown in Fig. \ref{fig:clsdstnem}.  The new estimates
slightly improve over the previous ones, suggesting once more that the leading source of error
in the present approach, as well in all previous, has to be found in the difficulty of properly handling the
term successive to the second in the virial expansion.

\subsection{Phase Coexistence}
\begin{figure}[tbh]
\vskip 1cm
\begin{center}
\includegraphics[width=.49\textwidth]{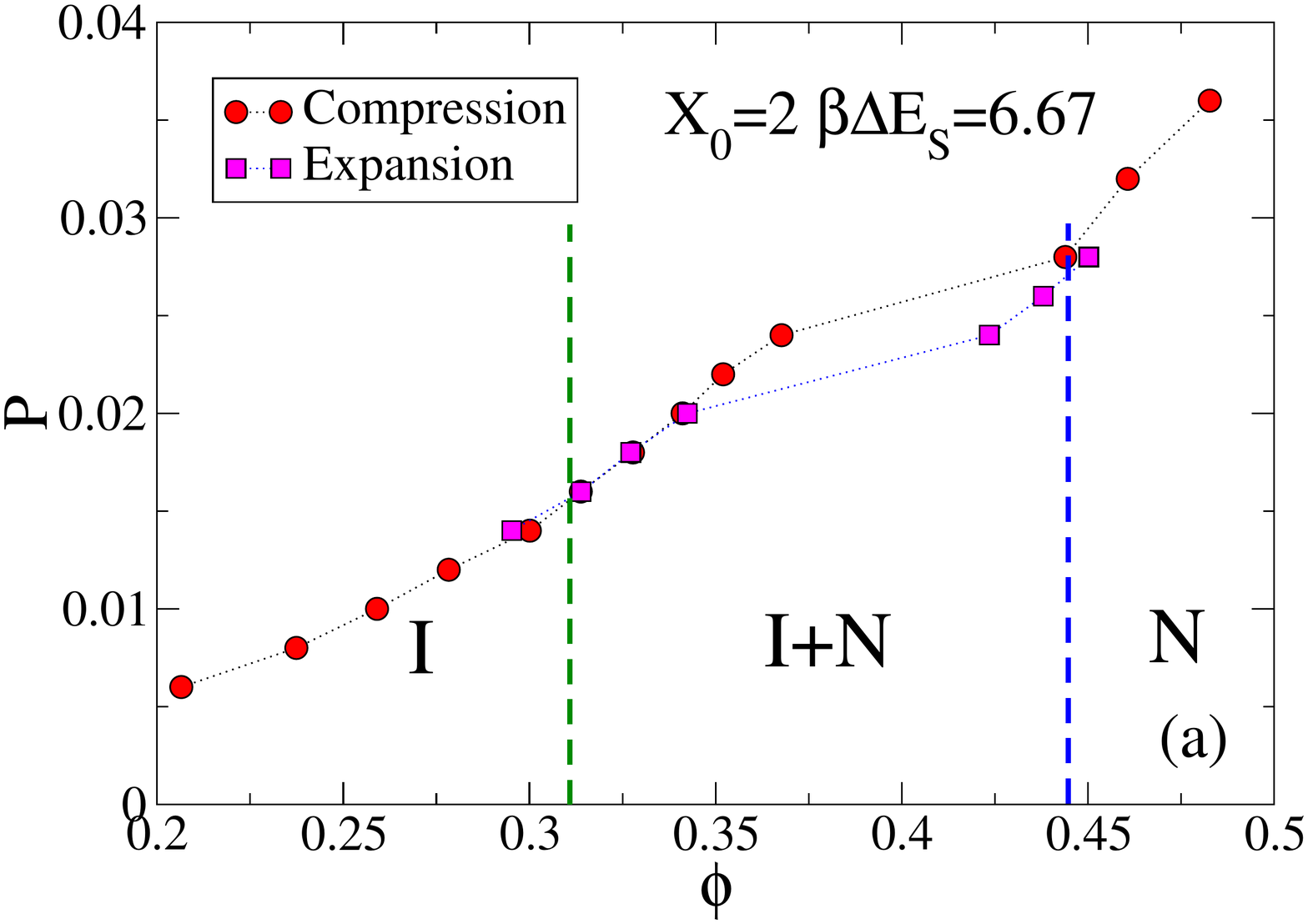}
\includegraphics[width=.49\textwidth]{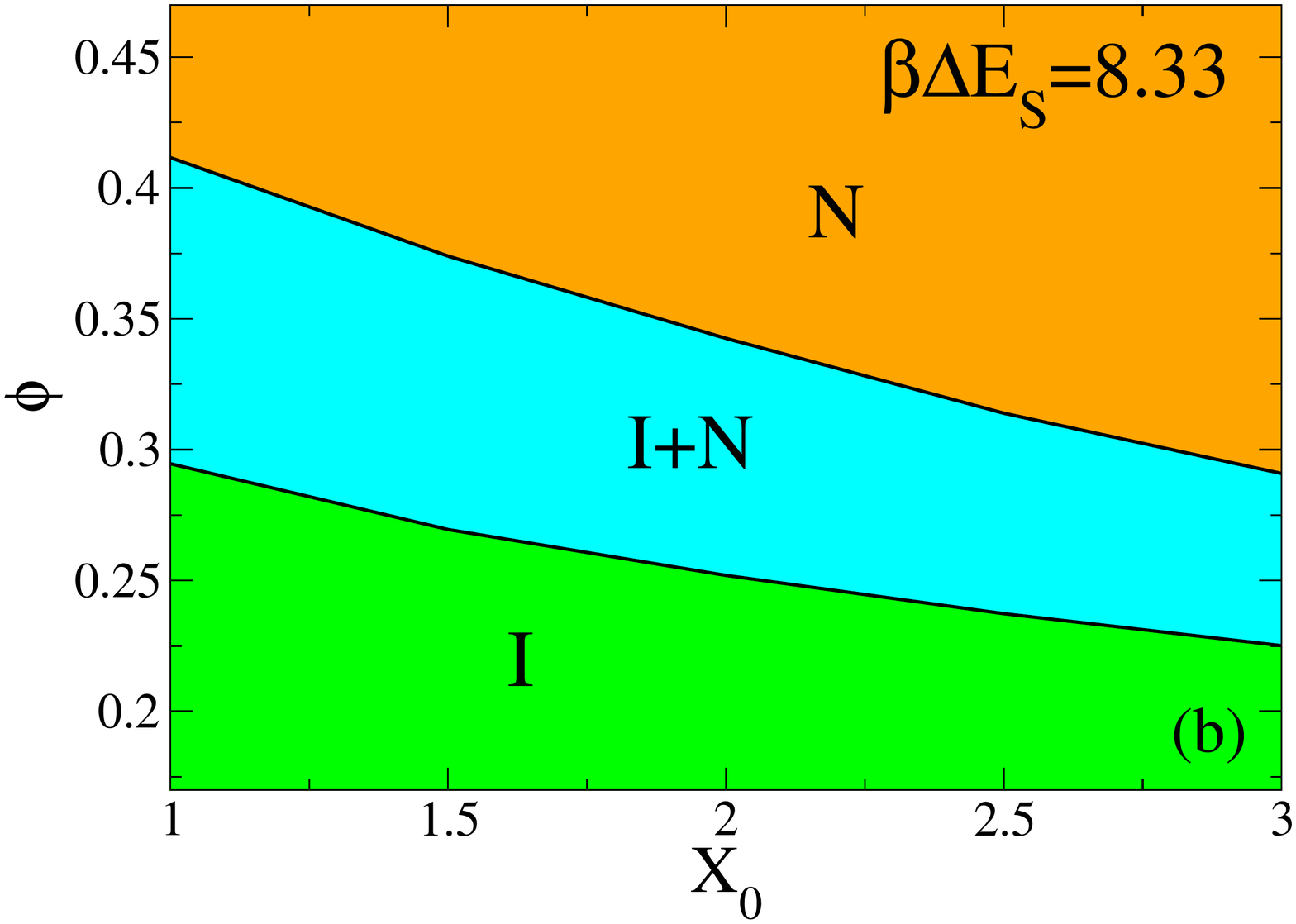}
\includegraphics[width=.49\textwidth]{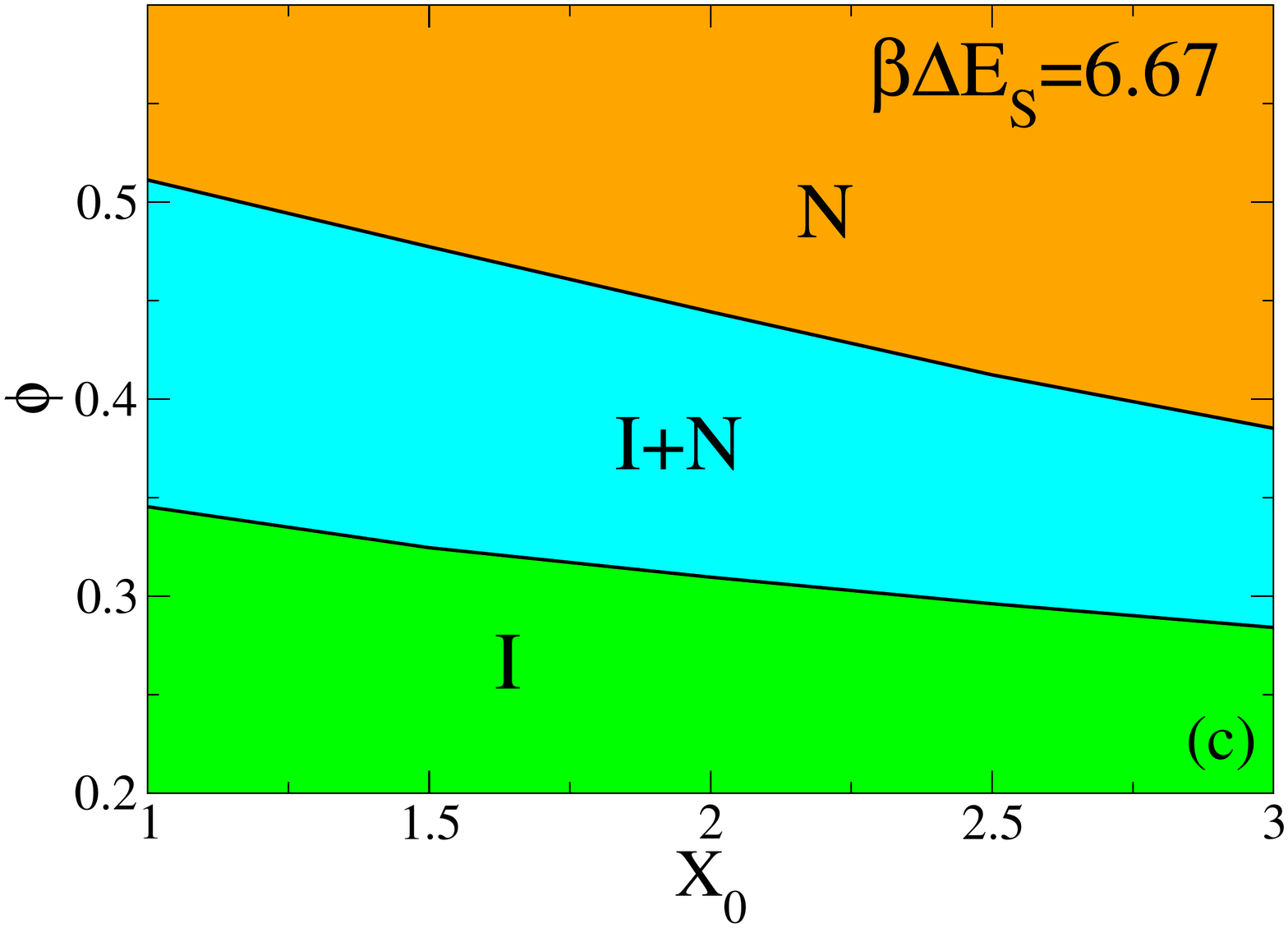}
\includegraphics[width=.49\textwidth]{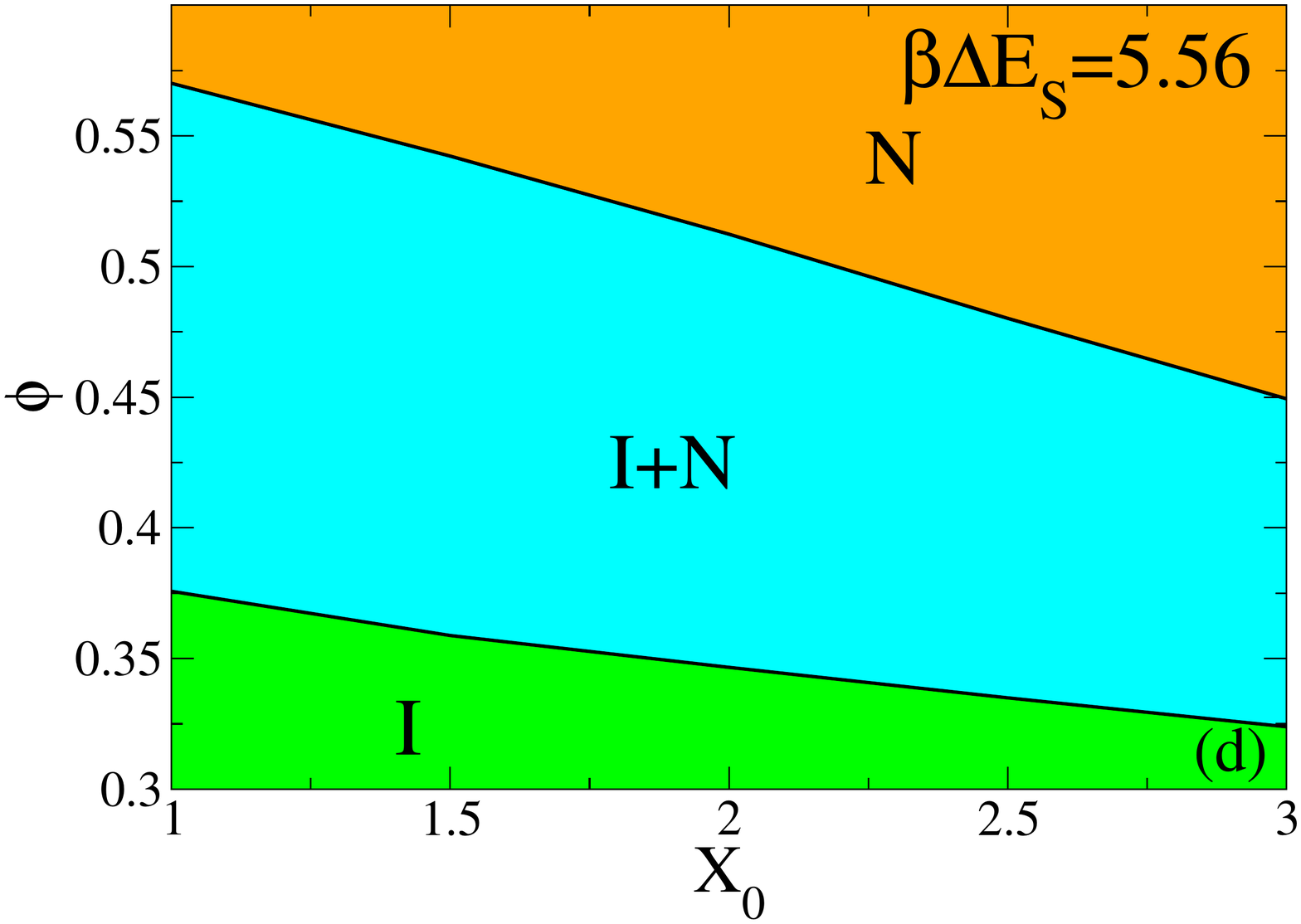}
\caption{ (a) Equation of state ($P$ vs $\phi$) calculated 
compressing an isotropic initial configuration (squares) or expanding an initial nematic configurations (circles).
Vertical dashed lines show the theoretical predictions for the phase boundaries. (b)-(d) Coexistence regions
predicted from theoretical calculations for the $3$  stacking energies values $\beta \Delta E_S$ investigated.}
\label{fig:PvsPhi}
\end{center}
\end{figure}

NPT-MC simulations provide a rough estimate of the location of 
 phase boundaries, being affected by the hysteresis associated to the metastability of the coexisting phases.
It is possible thus only to bracket the region of coexistence,  by selecting the first isotropic state
point on expansion runs which started from a nematic configuration and the first nematic state point on compression runs started from an isotropic configuration.  
We performed  NPT-MC simulations for $X_0=2$  and $\beta \Delta E_S=6.67$ 
in a wide range of pressures $P$ for a system of $1000$ SQs.
The resulting equation of state 
is shown in Fig. \ref{fig:PvsPhi}(a). As expected a clear hysteresis is observed, which allows us to
detect some overestimated boundaries for the isotropic-nematic transition.   The same figure also
reports the theoretical estimates of the transition.  The theoretical critical pressure is
smaller than the numerical one, resulting into a extended region of coexistence then numerically observed.
Comparing the values of the pressure predicted by the theory with the simulation values, we notice that
the main error arises from the pressure of the nematic phase which is underestimated.  
Finally,  Fig. \ref{fig:PvsPhi} (b)-(d) 
show  the predicted phase diagram for several values of  $\beta \Delta E_S$  as a function of the
elongation.  On increasing $\beta \Delta E_S$ (i.e. decreasing $T$ or increasing the
stacking energy)   there is a small decrease of  $\phi_I$  and a significant decrease of $\phi_N$, resulting
in an overall decrease of the $I-N$ coexistence region.  Such trend can be understood in term of
increase of the average chain length resulting from the increase of $\beta \Delta E_S$.
As expected, both  $\phi_I$ and  $\phi_N$ decrease on increasing $X_0$.

Finally we recall that in our model the persistence length $l_p$ depends on the elongation as discussed in Section \ref{sec:perslen}, anyway
the dependence on $l_p$ of the theoretical phase diagrams showed in Fig. \ref{fig:PvsPhi} (b)-(d) is negligible at the present level of accuracy 
of our theoretical calculations.  
\section{Comparison with experiments}
\label{sec:compwithexp}

Refs. \cite{BelliniScience07} and \cite{BelliniPNAS2010} report the critical concentrations ($c$), in $mg/ml$, for the $I-N$  transition of blunt-ended DNAD. 
These experimental data can be transformed into volume fractions once the relevant properties of DNAD are known
(DNAD molecular weight $m_{D}=660\,N_b$,  diameter $D \approx$ $2\; nm$,  length $L = N_b/3\; nm$, where $N_b$ is the number of bases in the sequence). 
The number density $\rho$ of DNADs is related to the  mass concentration
\begin{equation} 
\rho = \frac{c}{N m_{D}} 
\end{equation}  
Since $v_{d}=L D^2 \pi / 4 $ is the volume of a DNAD,  the volume fraction can be expressed as:
\begin{equation} 
\phi = \rho v_{d} = \frac{c L D^2 \pi}{4 N m_{D}} 
\label{eq:c2phi} 
\end{equation}  

\begin{figure}[tbh]
\includegraphics[width=.48\textwidth]{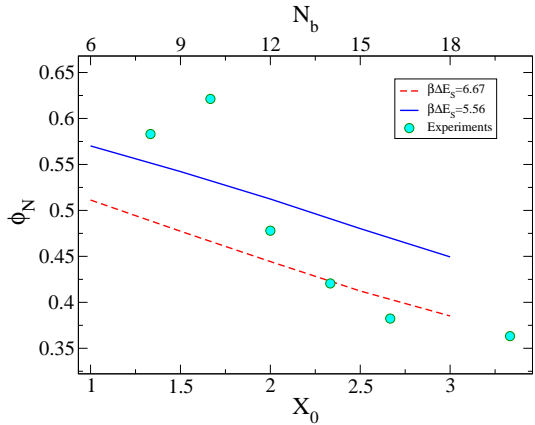}
\caption{Critical volume fractions $\Phi_N$ as a function of elongation $X_0$ (or equivalently $N_b$) from theoretical calculations  (for $\beta \Delta E_S=6.67$ and $\beta \Delta E_S=5.56$) and experiments\cite{BelliniScience07} (circles). }.
\label{fig:compexp}
\end{figure}

Data in Refs. \cite{BelliniScience07} and \cite{BelliniPNAS2010} put in evidence that blunt-end duplexes of equal length but different sequences may have different transition concentrations. As discussed in Ref. \cite{BelliniPNAS2010}
this phenomenon can be attributed to the slight differences in B-DNA helical conformation resulting from the difference in sequences.
These differences induce some curvature in the DNAD aggregates, in turn enhancing the transition concentration.
Indeed, sequences that are known to form straight double helices order into the $N$ phase at lower concentrations. 
Therefore, for each oligomer length in the range $8$-$16$ bases, we selected the lowest transition concentration among the ones experimentally determined, since these would be relative to duplexes closest to the symmetric 
monomers considered in the model. Such values have been reported in Figure \ref{fig:compexp} as a function of base number $N_b$ (top axis) and as a function of $X_0$ (bottom axis).
Apart for $N_b = 12$, of which a large number of sequences have been studied, the transition concentrations for the other $N_b$ values 
would probably be corrected to lower values if a larger number of sequences were experimentally explored. We would expect this to be particularly true for the shortest sequences, in which the effect of bent helices could be more relevant. 

In the experiments, DNADs are in a water solution with counter-ions resulting from the dissociation of the ionic groups of the phosphate-sugar chain. 
Given the high DNA concentration necessary for the formation of the $N$ phase, corresponding to concentration of nucleobases in the $1 M$ range, the ionic strength simply provided by the natural counter-ions is large enough to effectively screen electrostatic interactions between DNADs.  This becomes less true for the longest studied sequences, for which the transition concentration is lower. 
We hence decided to perform a small number of test experiments on the $N_b = 20$ oligomers with a double purpose: (i) determine more accurately the transition concentration value for this compound and (ii) test the effect of varying the ionic strength predicted by the model here described.
With respect to a fully screened DNAD where electrostatic repulsion can be neglected, a partly screened DNAD has a larger 
effective volume, thus filling a larger volume fraction of the solution, and a smaller axial ratio $X_0$, since electrostatic repulsion is equal in all directions.  Therefore, adding salt would bring about two competing effects: the reduction in particle volume enhances the 
concentration needed to reach the $I-N$ phase boundary,  while the grown of $X_0$ could favor the nematic ordering even at lower concentrations. 

In particular according to Eq. (\ref{eq:c2phi}) the following relation between the critical concentration $c_N$ and the
critical volume fraction $\phi_N$ holds:
\begin{equation}
c_N = \phi_N(X_0) \frac{4 N m_D}{L D^2}
\label{eq:cNphiN}
\end{equation}  
On the basis of the phase diagrams of Figure \ref{fig:PvsPhi} (b)-(d) $\phi_N(X_0)$ depends weakly on $X_0=L/D$, i.e. $\phi_N(X_0)\approx \phi_N^0$, where $\phi_N^0$ is a constant.
Hence the theory introduced in the present paper predicts that a reduction of DNAD effective volume due to the addition of salt (i.e. a decrease of $L D^2$ in Eq.~(\ref{eq:cNphiN}) leads to an overall increase of the concentration required for $N$ ordering.
 
We have measured the transition concentration of the self-complementary $20$mer $CGCGAAAATTTTCGCG$, 
a sequence whose $I-N$ transition at room temperature was previously measured and determined to be $c_{IN} \approx$ 200 $mg/ml$ \cite{BelliniScience07}. 
With the same method, based on the measurement of the refractive index of the solution, we determined the $I-N$ transition concentration at room temperature at three different ionic strengths. 
The values we obtained are $c_{IN} \approx 215\, mg/ml$ (no added salt), $c_{IN} \approx 320\, mg/ml$ ($0.8$ M NaCl), $c_{IN} \approx 380\,  mg/ml$ ($1.2$ M NaCl). The data indicate that the onset of the nematic ordering in solutions of $20$mers is indeed sensitive to the ionic strength, and that the transition concentration grows upon increasing the amount of salt,  as expected on the basis of our theoretical calculations for the present model. 
In Figure \ref{fig:compexp} we display the transition volume fraction derived by the transition concentration measured for $1.2$ M NaCl. At this ionic strength, the total concentration of Na$^+$ (dissociated from the oligomers + added with the salt) is about the same as the one resulting from counterions dissociated oligomers in the more concentrated solutions of shorter ($8$-$12$ mers) oligomers. 

Figure \ref{fig:compexp} compares the experimentally determined transition volume fractions 
with the values calculated from the model for  $\beta \Delta E_S=6.67$ and $\beta\Delta E_S=5.56$. 
Although experimental data are noisy, they fall in the range  $\Delta E_S\approx 5 - 7$ (in units of $k_B T$). 
Despite all the simplifying assumptions and despite the experimental uncertainty, results in Figure \ref{fig:compexp} provide a reasonable description of the $X_0$ dependence of $\phi_N$.

In comparing the model with the experimental results, it is necessary to take note of the fact that the stacking energy  between nucleobases, and thus the interaction energy $\Delta E_S$ between DNAD, is temperature dependent, i.e. its entropic component is relevant \cite{BelliniReview2011}. 
This is a general property of solvation energies and thus it is in line with the notion that stacking forces are mainly of hydrophobic nature. 
Therefore, the range of values for $\Delta E_S$ determined in Figure \ref{fig:compexp}
should be compared to the values of $\Delta G$ for the stacking interactions at the temperature at which the experiments were performed. 
Overall, the estimate of $\Delta E_S$ here obtained appears as in reasonable agreement with the free energies
involved in the thermodynamic stability of the DNA double helices and confirms the rough estimate 
that was given before (see the supporting online material associated to Ref.~\cite{BelliniScience07}).


\section{Conclusions}
\label{sec:concl}

In this study we have developed a free energy functional in order to calculate the phase diagram of 
bi-functional quasi-cylindrical monomers, aggregating into equilibrium chains,  with respect to isotropic-nematic transition.  The model has been inspired by experiments on the aggregation of short DNA which exhibit at sufficiently high concentration nematic phases and the comparison between the theoretical predictions and the experimental results allows us to provide an estimate of the stacking energy, consistent with previous propositions.

Our approach is  quite general, parameter free and not restricted to particular shapes. We provide techniques to
evaluate the bonding volume and  the excluded volume, which enters into our formalism via the
Parson-Lee decoupling approximation.   We build on previous work, retaining the
discrete cluster size description of Ref.~\cite{KindtJCP04} and the Parson-Lee factor for the excluded volume contribution proposed in Ref.~\cite{GlaserMC}. With respect to previous approaches we (i) explicitly account for the
entropic and energetic contributions associated to bond formation; (ii) we do not retain any adjustable fit parameter.

The  resulting description of  the isotropic phase is rather satisfactory and quantitative up to $\phi \approx 0.2$.
The description of the nematic phase partially suffers from some of the
approximations made in deriving the free energy functional. More specifically, 
several signatures point toward the failure of the Parsons decoupling approximation in the 
$\phi$ range typical  of the nematic phase.   While there is a sufficient understanding of the quality of
such approximation for monodisperse object \cite{LekkMolPhys09,CinacchiJCP04,SpheroCylRec,LekkJPCB01,AllenJCP96,CuetosPRE07}, work need to be done to assess the origin of the failure of this approximation in the equilibrium polymer case and to propose improvements.

We finally remind that the model here introduced does not consider the azimuthal rotations of each monomer around its axis. This neglect is adequate when the aggregation does not entail constraints in the azimuthal freedom of the monomers. This is the case of base stacking, in which the angular dependence of the stacking energy is arguably rather small. However, this is not the case of DNAD interacting through the pairing of overhangs and of the LC ordering of RNA duplexes.  Because of its A-DNA-type structure, the terminal paired bases of RNA duplexes are significantly tilted with respect to the duplex axis, thus establishing even in the case of
blunt-ended duplexes a link between the azimuthal angle of the aggregating duplexes and the straightness of the aggregate. 
However, with minor modifications the model here introduced could become suitable to include these additional situations. The limiting factor in developing such extension is the lack of knowledge to quantify the azimuthal constraints implied by these interactions. This situation, as well as the effects of off-axis components of the end-to-end interduplex interactions, will be explored in a future work.

\section{Acknowledgments}
CDM and FS acknowledge support from ERC (226207-PATCHYCOLLOIDS). 

\section{Appendix A}
\label{AppendixA}
Here we provide a justification for the use of Parsons decoupling approximation in the case of linear chains poly-disperse in length
(with distribution $\nu(l)$), 
based on the extension of Onsager's second-virial theory to mixtures of non-spherical hard bodies proposed in Ref. \cite{JacksonJCP08}.
 The contribution $F_{excl}$ to the free energy due to excluded volume interactions between chains can be written
 if we neglect intrachain interactions \cite{HonnellJCP87,JacksonJCP08}: 
\begin{eqnarray}
\frac{\beta F_{excl}}{V} &=& \frac{\rho}{6} \int_0^{\rho} d\rho' \int d{\bf r} \int d{\bf\Omega}_1 d{\bf\Omega}_2 \sum_{ll'} \frac{\nu(l)\nu(l')}{{\rho}^2}\nonumber\\
&& g_{ll'}({\bf r}, {\bf \Omega}_1,{\bf \Omega}_2) f({\bf \Omega}_1) f({\bf \Omega}_2)\, {\bf r}\cdot \nabla_{\bf r}V_{HC}({\bf r}, {\bf \Omega}_1,{\bf \Omega}_2) 
\label{eq:helmsplit}
\end{eqnarray}
where ${\bf r}$ is the distance between the centers of mass of the two chains $1$ and $2$,  ${\bf \Omega}_1=\{{\bf u}_1^1,\ldots {\bf u}_l^1\}$ and ${\bf \Omega}_2=\{{\bf u}_1^2,\ldots {\bf u}_{l'}^2\}$ are the orientations of the two chains, where ${\bf u}_i^\alpha$ is the orientation
of monomer $i$ belonging to chain $\alpha=1,2$, $g_{ll'}({\bf r}, {\bf \Omega}_1,{\bf \Omega}_2)$ is the molecular radial distribution function of
the mixture,  which represents the correlations between two chains of length $l$ and $l'$, whose relative distance is ${\bf r}$ and
which have orientations  ${\bf \Omega}_1$ and ${\bf \Omega}_2$ respectively, $V_{HC}({\bf r}, {\bf \Omega}_1,{\bf\Omega}_2)$ is the hard-core part of the interaction potential and $f({\bf \Omega}_\alpha)$ is the angular distribution function of chain $\alpha$.
We note that in Eq. (\ref{eq:helmsplit}) the integration in $\rho'$ is performed keeping fixed all the parameters related to $f({\bf\Omega}_\alpha)$.
Neglecting intra-chain interactions is equivalent to ignore self-overlaps of chains, an assumption which is appropriate if chain
length is not much greater than its persistence length and the chains can be considered non-extensible.
 

Parsons decoupling approximations in this case accounts to putting:
\begin{equation}
g_{ll'}({\bf r}, {\bf \Omega}_1,{\bf \Omega}_2) = g^{HS}_{ll'}[r/\sigma_{l l'}(\hat r, {\bf \Omega}_1,{\bf \Omega}_2)]
\label{eq:PDA}
\end{equation} 
where $g^{HS}_{ll'}$ is the radial distribution function of a mixture of hard spheres and $\sigma_{ll'}(\hat r, {\bf \Omega}_1,{\bf \Omega}_2)$ is an angle-dependent range parameter which depends on chain lengths $l$ and $l'$.
If the pair interaction is of the special form
\begin{equation}
V_{HC}({\bf r}, {\bf\Omega}_1,{\bf \Omega}_2) = V_{HC}[r/\sigma_{l l'}(\hat r, {\bf \Omega}_1, {\bf \Omega}_2)]
\label{eq:VHCapprox}
\end{equation}
noting that ${\bf r }\cdot\nabla_{\bf r} = r \frac{\partial}{\partial r}$,  Eq. (\ref{eq:helmsplit}) becomes:
\begin{eqnarray}
\frac{\beta F_{excl}}{V} &=& \frac{\rho}{6} \int_0^{\rho} d\rho' \int d\hat  r\, d{\bf\Omega}_1\, d{\bf \Omega}_2 \sum_{ll'} \frac{\nu(l)\nu(l')}{{\rho'}^2}\nonumber\\
&&\int dr\; r^3 g^{HS}_{ll'}(r/\sigma_{ll'})f({\bf \Omega}_1) f({\bf \Omega}_2) \, \frac{\partial\, V_{HC}(r/\sigma_{ll'})}{\partial r} 
\label{eq:Fdecl}
\end{eqnarray} 
With the substitution $y=r/\sigma_{ll'}$  from Eq. (\ref{eq:Fdecl}) one obtains:
\begin{eqnarray}
\frac{\beta F_{excl}}{V} &=& \frac{\rho}{2}\sum_{ll'} \frac{1}{3}\frac{\nu(l)\nu(l')}{{\rho}^2} 
\int_0^{\rho} d\rho'\int d\hat r \,d{\bf\Omega}_1\, d{\bf \Omega}_2
 \nonumber\\
 &&\int dy\; y^3 \frac{\partial V_{HC}(y)}{\partial y} g^{HS}_{ll'}(y) f({\bf \Omega}_1) f({\bf \Omega}_2) \sigma_{ll'}^3   
\label{eq:FexclRedY}
\end{eqnarray}

The derivative of $V_{HC}$ is a delta function hence we need only to evaluate the value of $g^{HS}_{ll'}(y)$ at contact (i.e. $y=1^+$) 
and Eq. (\ref{eq:FexclRedY}) becomes:
\begin{eqnarray}
\frac{\beta F_{excl}}{V} &=& \frac{\rho}{2} \sum_{ll'}  \frac{\nu(l)\nu(l')}{{\rho}^2}  \int d\rho' g_{ll'}^{HS}(1^+)
 \int d\hat r \,d{\bf \Omega}_1\, d{\bf \Omega}_2\nonumber  f({\bf \Omega}_1) f({\bf \Omega}_2) \frac{\sigma_{ll'}^3}{3} 
\end{eqnarray} 
This expression tends to Parson's expression when the system is monodisperse ($\nu(l)=\rho \delta_{l,1}$).
In the specific case of spherical particles
$\sigma_{ll'}(\hat r, {\bf \Omega}_1, {\bf \Omega}_2)=  \sigma(\hat r, {\bf \Omega}_1, {\bf \Omega}_2) =      \sigma $
and
\begin{equation}
\sum_{ll'} \nu(l)\nu(l')  f({\bf \Omega}_1) f({\bf \Omega}_2) \int d\hat r \,d{\bf\Omega}_1\, d{\bf \Omega}_2 \frac{1}{3} \sigma (\hat r, {\bf \Omega}_1, {\bf \Omega}_2) = \frac{4 \pi}{3} \sigma^3
\end{equation}
i.e. the excluded volume of two spheres of diameter $\sigma$.  Hence
we are allowed to make the identification:
\begin{eqnarray}
v_{excl}(l, l') &=&  \int d\hat r \,d{\bf\Omega}_1\, d{\bf \Omega}_2 \frac{1}{3}  f({\bf \Omega}_1) f({\bf \Omega}_2) \sigma_{ll'}^3(\hat r, {\bf \Omega}_1, {\bf \Omega}_2)
\label{eq:appAvexcl}
\end{eqnarray}
and write:
\begin{equation}
\frac{\beta F_{excl}}{V} = \frac{\rho}{2} \sum_{ll'}  \frac{\nu(l)\nu(l')}{{\rho}^2} \left [ \int d\rho' g_{ll'}^{HS}(1^+)  \right ]  v_{excl}(l, l') 
\label{eq:fexclPDA}
\end{equation}
We note that the identification made in Eq. (\ref{eq:appAvexcl}) can be also further justified using the same reasonings given in Sec. \ref{Sec:theory}. 
As discussed in Ref. \cite{JacksonJCP08} a possible expression for $g_{ll'}^{HS}$ is the one derived by Boubl\'ik \cite{BoublikJCP70}, which
generalizes the Carnahan--Starling relation \cite{CS_69} for pure hard spheres to the case of mixtures, i.e.
\begin{equation}
g^{HS}_{ll'} (1^+) = \frac{1}{1-\zeta_3} + \frac{3\,\zeta_2}{(1-\zeta_3)^2}\frac{\hat\sigma_{ll} \hat\sigma_{l'l'}}{\hat\sigma_{ll}+\hat\sigma_{l'l'}} + 
\frac{2\, \zeta_2^2}{(1-\zeta_3)^3} \frac{\left ( \hat\sigma_{ll} \hat\sigma_{l'l'} \right )^2}{(\hat\sigma_{ll}+\hat\sigma_{l'l'})^2}
\label{eq:ghsmix}
\end{equation}
where $\hat\sigma_{ll}$ is the diameter of an hard sphere corresponding to a chain of length $l$ and 
$\zeta_n = (\pi / 6)\sum_l \nu(l) \hat\sigma_{ll}^n$.
To map the system of polydisperse chains onto the equivalent mixture of hard spheres we need an expression for $\hat\sigma_{ll}$. 
According to Ref. \cite{JacksonJCP08}, the simplest choice is to consider spheres having the same volume of the corresponding linear chain of length $l$, i.e.
\begin{equation}
v_d = \frac{\pi}{6\, l} \hat\sigma_{ll}^3
\label{eq:sphvol}
\end{equation}
where we recall that $v_d$ is the volume of a monomer.
Although in principle we could use  Eq. (\ref{eq:fexclPDA}) together with Eqs. (\ref{eq:ghsmix}) and (\ref{eq:sphvol})
to calculate the free energy contribution due to the excluded volume between particles, if we make the further assumption that
 \begin{equation}
 g_{ll'}^{HS}(1^+) \approx g^{HS}(1^+)
 \label{eq:monoapprox}
 \end{equation}
i.e. if we approximate the radial distribution function of the hard spheres mixture at contact with that of a monodisperse system of hard spheres having the same total volume fraction (i.e. setting in Eq. (\ref{eq:ghsmix}) $\hat\sigma_{ll'}=\hat\sigma$ with $M v_d = (\pi/6) \sigma^3$), 
we finally obtain 
\begin{equation}
\frac{\beta F_{excl}}{V} = \frac{\eta(\phi)}{2} \sum_{ll'} \nu(l)\nu(l')  v_{excl}(l, l') 
\label{eq:fexclPDAeta}
\end{equation}
where we used the Carnahan-Starling expression for  $g^{HS}(1^+;\rho')$ and we performed the integration in $\rho'$.
Eq. (\ref{eq:fexclPDAeta}) is exactly the expression for the contribution to the free energy due to steric repulsion 
which we used in Section \ref{Sec:theory}.
In summary according to the above derivation we argue that  Eq. (\ref{eq:fexclPDAeta}) can be not accurate 
at high volume fractions due to the approximations made in Eqs. (\ref{eq:PDA}) (i.e. the Parsons decoupling approximation)
and (\ref{eq:monoapprox}).
Within the present treatment Eq. (\ref{eq:fexclPDAeta}) is also not appropriate 
for chains with $l \gg l_p$ because, as already noted, chain self-overlaps can be significant and 
the hard body pair potential $V_{HC}$ does not have the special form assumed in Eq.~(\ref{eq:VHCapprox}).
 
We finally note that the approximation made in Eq.~(\ref{eq:monoapprox}) can be avoided if one resorts to Eq. (\ref{eq:fexclPDA}) instead of Eq. (\ref{eq:fexclPDAeta}), although the required free energy calculations would become much more complicated. 
Anyway we verified for the isotropic phase that employing Eq.~(\ref{eq:fexclPDA}) instead of Eq.~(\ref{eq:monoapprox}) does not 
provide any appreciable improvement in the present case.
\section{Appendix B}
\label{AppendixB}
The procedure to calculate the excluded volume $v_{excl}$ in the isotropic phase consists in performing $N_{att}$ attempts of inserting two chains of length $l$ in a box of volume $V$ as described in the following:
\begin{enumerate}
\item Set the counter $N_{ov}=0$
\item Build first chain of length $l$ randomly, according to the following procedure:
\begin{enumerate}
\item Insert a first randomly oriented monomer.
\item Insert  a monomer $\cal M$ bonded to a 
free site $\cal S$ on chain ends ($\cal S$ can be chosen randomly among the two free sites of the partial chain). The orientation of $M$ will be random and its position will be chosen randomly within the available bonding volume between $\cal M$ and $\cal S$. The bonding volume between $\cal M$ and $\cal S$ is defined as the volume corresponding to all possible center of mass positions of $\cal M$ with $\cal M$ bonded to $\cal S$. 
\item If the number of monomer inserted is $l$ terminate otherwise go to 1).
\end{enumerate}
where the first monomer inserted is placed in the center of the box
and it is oriented with its attractive sites parallel to the $x$-axis.
\item Build a second chain of length $l$, where the first monomer inserted is placed randomly within the 
simulation box with a random orientation.
\item Increase $N_{ov}$ by $1$ if two monomers belonging to different chains overlap and 
the two chains are either not self-overlapping or forming a closed loop.
\item if the number of attempts is less than $N_{att}$ go to 2) otherwise terminate.
\end{enumerate}
Then $v_{excl}$ can be calculated as follows:
\begin{equation}
v_{excl} = \frac{N_{ov}}{N_{att}} V
\label{eq:numcov}
\end{equation} 
A reasonable choice for the total number of attempts is $N_{att}=10^6$.
In a similar fashion one can also calculate the bonding volume \cite{BianchiJCP07} 
between two monomers.  In this case one monomer is kept fixed in the center of the simulation box and 
the other one is inserted with random position and orientation for a total of $N_{att}$ attempts.
The bonding volume will be:
\begin{equation}
V_b = \frac{N_{bond}}{4 N_{att}} V
\label{eq:Vbnum} 
\end{equation}
where the factor $4$ accounts for the fact that two particles can form $4$ different possible bonds
and $N_{bond}$ is the number of times that the two monomers were bonded after a random insertion.
Finally with the same procedure used to calculate the excluded volume in the isotropic phase
we can evaluate the excluded volume in the nematic phase. The only difference is that now
monomers have to be inserted with an orientation extracted from the Onsager angular distribution defined in Eq.  (\ref{eq:fons}),
so that the excluded volume depends also on the parameter $\alpha$.
Again if $N_{ov}$ is the number of times that two monomers belonging to different clusters overlap 
and $N_{att}$ is the total number of attempts then we have:
\begin{equation}
v_{excl}(l,l,\alpha) = \frac{N_{ov}}{N_{att}} V
\label{eq:numcovalpha}
\end{equation}

\section{Appendix C}
\label{AppendixC}
\begin{figure}[tbh]
\vskip 0cm
\begin{center}
\includegraphics[width=.49\textwidth]{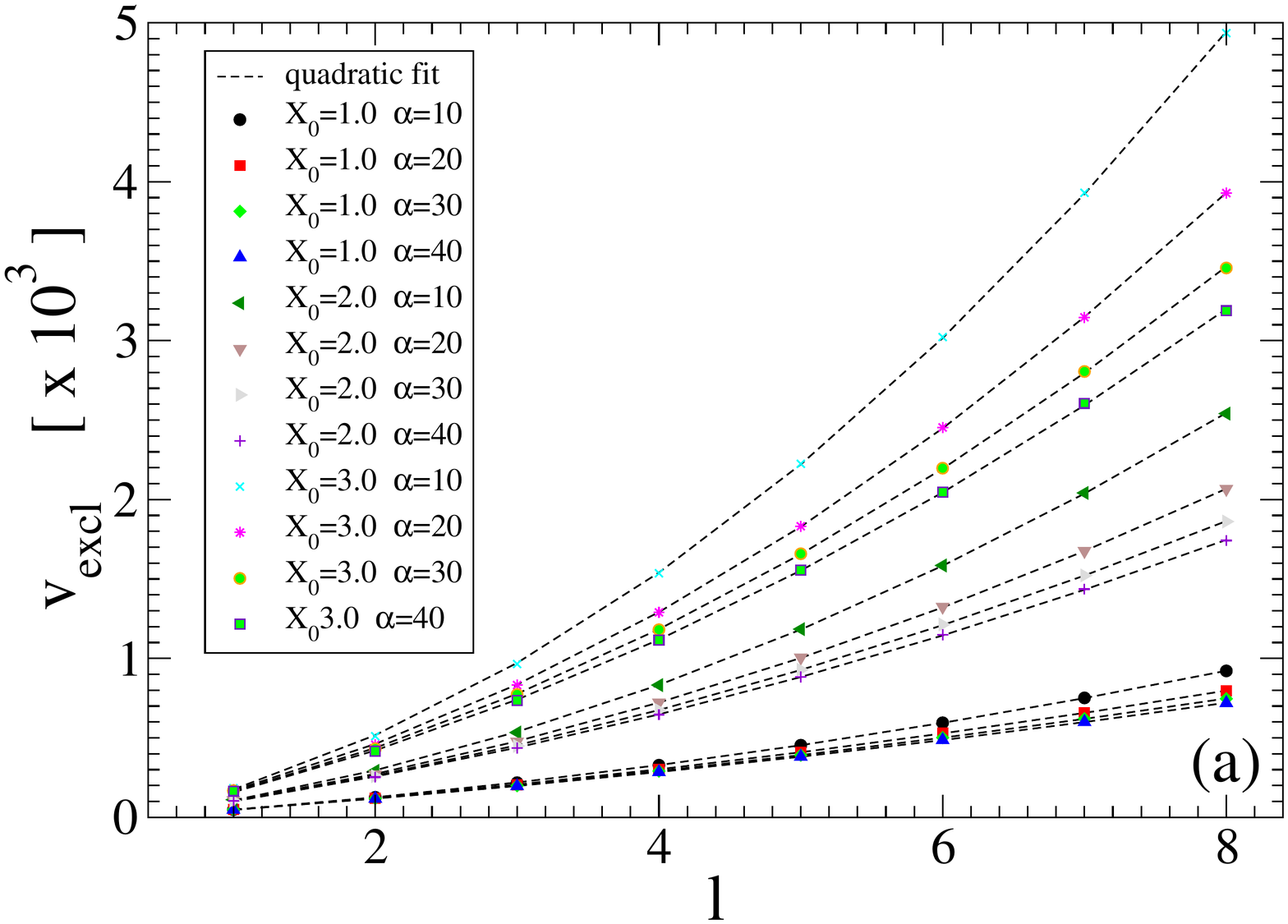}
\includegraphics[width=.49\textwidth]{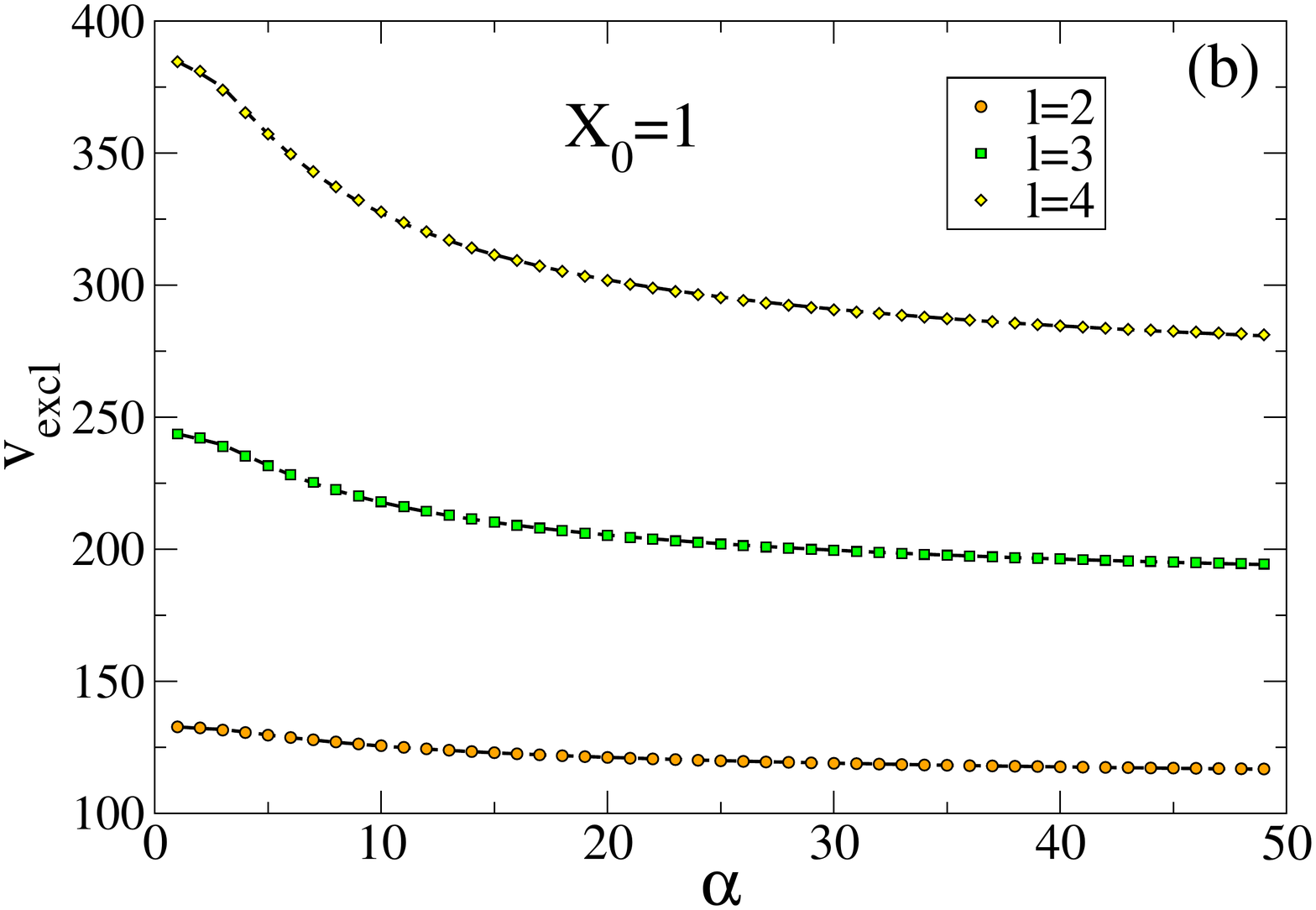}
\includegraphics[width=.49\textwidth]{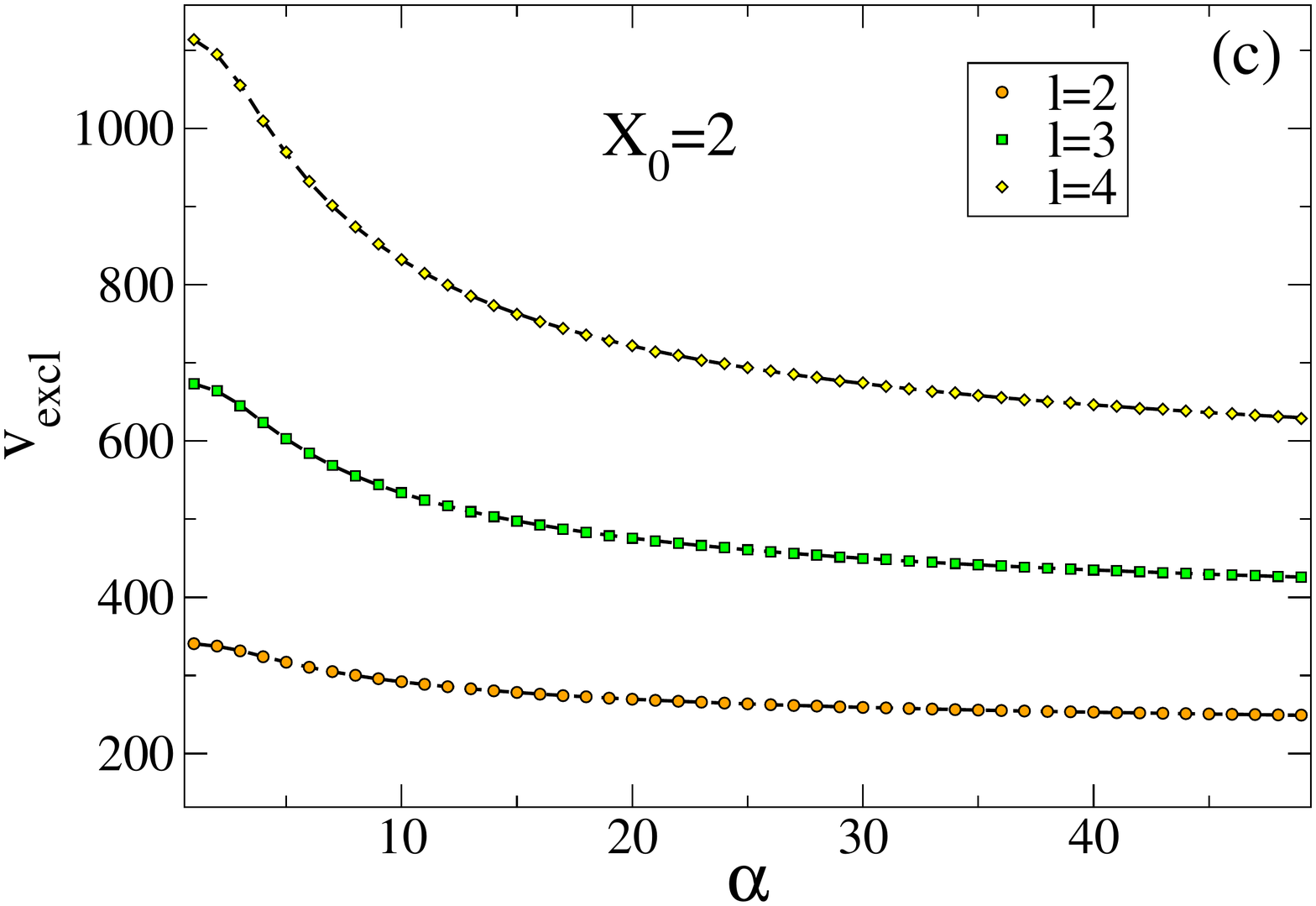}
\includegraphics[width=.49\textwidth]{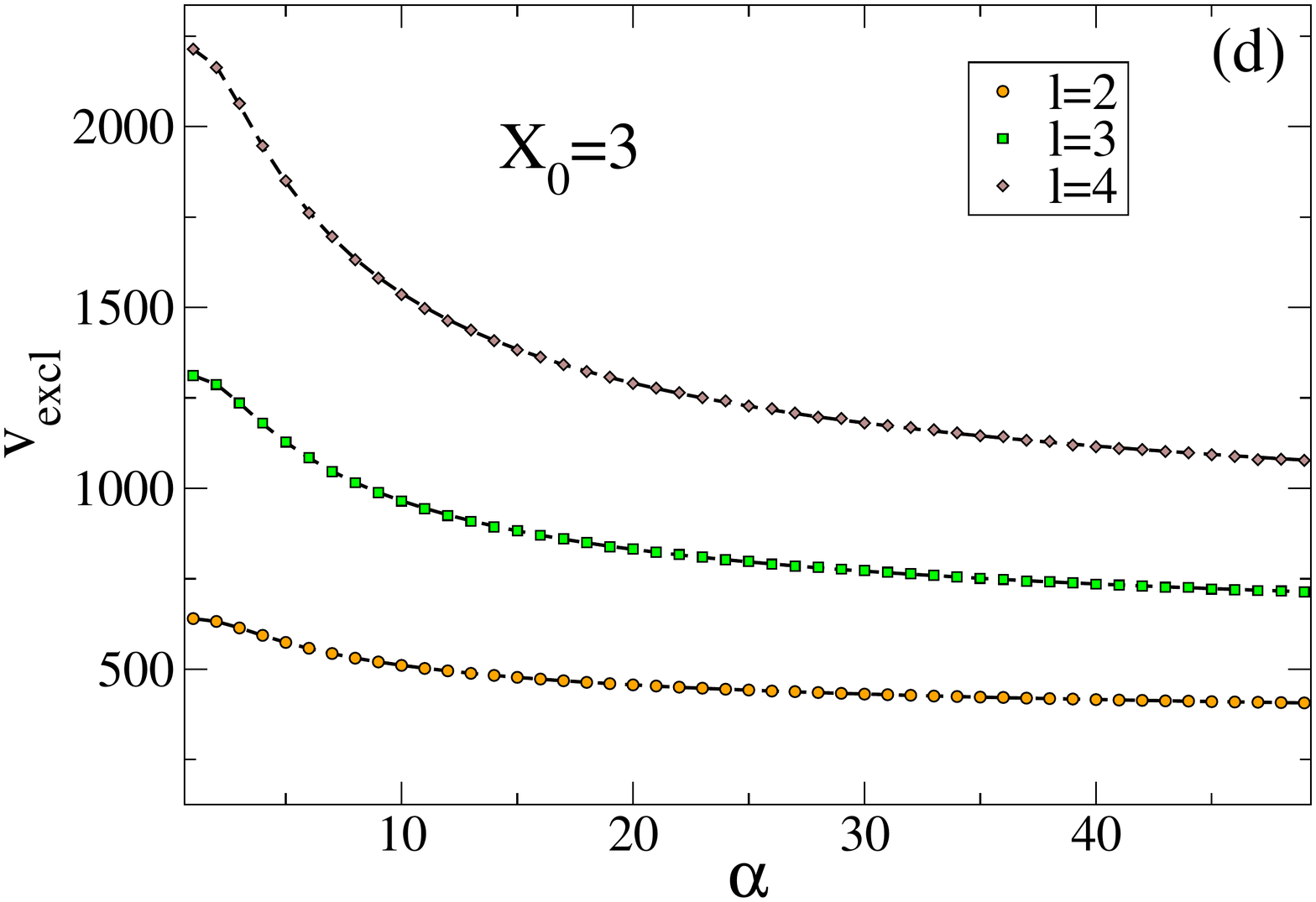}
\caption{ (a)  Excluded volume of two chains of length $l$ as a function of chain length  for the nematic cases $\alpha=10,20,30,40$ and three different elongations
$X_0=1,2,3$.
(b)-(d) Excluded volume in the nematic phase calculated numerically as a function of $\alpha$ 
for two chains of equal length $l$, where $l=2,3,4$, composed of 
 monomers with $X_0=1,2,3$.}
\label{Fig:covalpha}
\end{center}
\end{figure}

In this Appendix we explain how to calculate the parameters $ A_N(\alpha)$, $k_N(\alpha)$ and $B_N(\alpha)$
of the nematic free energy functional.
As a preliminary step we check that $v_{excl}(l,l',\alpha)$ for a fixed value of $\alpha$ is a second order polynomial
of $l$ and $l'$ as assumed in Eq. (\ref{eq:genvexcl}). In Fig. \ref{Fig:covalpha} (a) we plot $v_{excl}(l,l,\alpha)$  as a function
of $l$ for different values of $\alpha$ and $X_0$, $v_{excl}(l,l,\alpha)$  can be well represented by
a parabolic function, in agreement with Eq.~(\ref{eq:genvexcl}). 

We start by  observing that the $\alpha$ dependence of $A_N(\alpha)$,  $k_N(\alpha) $ and $B_N(\alpha)$ in the case of hard cylinders following the Onsager distribution can be expanded in powers of $\alpha^{-1/2}$ as

\begin{eqnarray}
A_N(\alpha) &=&  c_{00} + \frac{c_{01}}{\alpha^{1/2}} + \frac{c_{02}}{\alpha} + \frac{c_{03}}{\alpha^{3/2}} + \frac{c_{04}}
{\alpha^2}\nonumber\\  
k_N(\alpha) &=&  c_{10} + \frac{c_{11}}{\alpha^{1/2}} + \frac{c_{12}}{\alpha} + \frac{c_{13}}{\alpha^{3/2}} + \frac{c_{14}}
{\alpha^2}\nonumber\\  
B_N(\alpha) &=&  c_{20} + \frac{c_{21}}{\alpha^{1/2}} + \frac{c_{22}}{\alpha} + \frac{c_{23}}{\alpha^{3/2}} + \frac{c_{24}}
{\alpha^2}
\label{eq:xi} 
\end{eqnarray}

where $c_{ij}$ are the elements of the $3\times4$  matrix $\bf C$. In the case of cylinders, some of the $c_{ij}$ vanishes\cite{Odijk86}.   We assume here that the same $\alpha$ dependence holds for SQ.



 
 In view of this result the co-volume as function of $l$ and $\alpha$ can be expressed as 
\begin{equation}
v_{excl}^{(fit)}(\alpha; X_0, l) = d_{l0} + \frac{d_{l1}}{\alpha^{1/2}} + \frac{d_{l2}}{\alpha} + 
\frac{d_{l3}}{\alpha^{3/2}} +  \frac{d_{l4}}{\alpha^{2}}
\label{eq:vexclfit} 
\end{equation}
where $d_{l,p}$, for $p=0,4$ are fitting parameters.
Fig.\ref{Fig:covalpha} (b)-(d) shows the numerical calculation of the covolume varying $\alpha$ for three particular elongations ($X_0=1, 2 ,3$), together with fits to the functional form of Eq. (\ref{eq:vexclfit}). 

The good quality of the fits (reduced $\chi^2$ is always much less than $1$ for all fits) suggests that
retaining terms up to $O(1/\alpha^2)$ is to the present level of accuracy of our calculations 
absolutely appropriate.

From these fits we can estimate 
the matrix $\bf C$
needed to evaluate the free energy in the nematic phase for each $X_0$.
If we define in fact the following matrix $\bf P$ and the vectors ${\bf q}_p$, with $p=0\ldots 4$ as follows:
\begin{equation}
{\bf P}=\left (\begin{matrix} 1 & l_a & l_a^2\\1 &  l_b & l_b^2\\1 &  l_c & l_c^2\end{matrix}\right )
\;\; {\bf q}_p = \left ( \begin{matrix}d_{l_a p} \\d_{l_b p}\\d_{l_c p} \end{matrix} \right )
\label{eq:PmatrixDef}
\end{equation}
where $l_a$, $l_b$ and $l_c$ are three different chain lengths for which we calculated the $v_{excl}$ as a function of $\alpha$, then we can calculate the matrix elements of $\bf C$ in the following way:
\begin{equation}
2 \left (\begin{matrix} c_{0p} \\v_d c_{1p} \\  X_0^2 c_{2p}\end{matrix} \right ) = 
 {\bf P}^{-1} {\bf q}_p.
\label{eq:calcnempars}
\end{equation}

\bibliography{cokecan}
\end{document}